\documentclass[a4paper,11pt]{article}
\usepackage{jheppub} % for details on the use of the package, please see the JHEP-author-manual

\usepackage[T1]{fontenc} % if needed
\usepackage[utf8]{inputenc}

\usepackage{bm}
\usepackage[strings]{underscore}

\def\empile#1\above#2{\mathrel{\mathop{\kern 0pt#1}\limits_{#2}}}

\newcommand{\non}{\nonumber\\}

\newcommand{\absol}[1]{\left| #1 \right|}
\newcommand{\TeV}{\,\hbox{TeV}}
\newcommand{\GeV}{\,\hbox{GeV}}
\newcommand{\MeV}{\,\hbox{MeV}}

\newcommand{\sll}{\raise.15ex\hbox{$/$}\kern-.43em\hbox{$l$}}
\newcommand{\slepsilon}{\raise.15ex\hbox{$/$}\kern-.53em\hbox{$\epsilon$}}
\newcommand{\slvarepsilon}{\raise.15ex\hbox{$/$}\kern-.53em\hbox{$\varepsilon$}}
\newcommand{\slL}{\raise.15ex\hbox{$/$}\kern-.53em\hbox{$L$}}
\newcommand{\slP}{\raise.15ex\hbox{$/$}\kern-.53em\hbox{$P$}}
\newcommand{\slp}{\raise.1ex\hbox{$/$}\kern-.63em\hbox{$p$}}
\newcommand{\slq}{\raise.1ex\hbox{$/$}\kern-.53em\hbox{$q$}}
\newcommand{\slv}{\raise.1ex\hbox{$/$}\kern-.63em\hbox{$v$}}
\newcommand{\slR}{\raise.15ex\hbox{$/$}\kern-.53em\hbox{$R$}}
\newcommand{\slQ}{\raise.15ex\hbox{$/$}\kern-.53em\hbox{$Q$}}
\newcommand{\slK}{\raise.15ex\hbox{$/$}\kern-.53em\hbox{$K$}}
\newcommand{\slk}{\raise.15ex\hbox{$/$}\kern-.53em\hbox{$k$}}
\newcommand{\slSigma}{\raise.15ex\hbox{$/$}\kern-.53em\hbox{$\Sigma$}}
\newcommand{\slcalP}{\raise.15ex\hbox{$/$}\kern-.63em\hbox{$\cal P$}}
\newcommand{\slA}{\raise.15ex\hbox{$/$}\kern-.73em\hbox{$A$}}
\newcommand{\slbfA}{\raise.15ex\hbox{$/$}\kern-.73em\hbox{${\imb A}$}}
\newcommand{\slpartial}{\raise.15ex\hbox{$/$}\kern-.53em\hbox{$\partial$}}
\newcommand{\sla}{\raise.15ex\hbox{$/$}\kern-.53em\hbox{$a$}}
\newcommand{\slb}{\raise.15ex\hbox{$/$}\kern-.53em\hbox{$b$}}
\newcommand{\slc}{\raise.15ex\hbox{$/$}\kern-.53em\hbox{$c$}}
\newcommand{\slD}{\raise.15ex\hbox{$/$}\kern-.53em\hbox{$D$}}
\newcommand{\slC}{\raise.15ex\hbox{$/$}\kern-.53em\hbox{$C$}}

\title{\boldmath  Comparison of improved TMD and CGC frameworks in forward quark dijet production}

\author[a]{Hirotsugu Fujii,}
\author[b]{Cyrille Marquet,}
\author[c,d,e]{and Kazuhiro Watanabe}

\affiliation[a]{Institute of Physics, University of Tokyo, Tokyo 153-8902, Japan}
\affiliation[b]{CPHT, CNRS, Ecole Polytechnique, Institut Polytechnique de Paris, 91128 Palaiseau, France}
\affiliation[c]{Theory Center, Jefferson Laboratory, Newport News, Virginia 23606, USA}
\affiliation[d]{Physics Department, Old Dominion University, Norfolk, Virginia 23529, USA}
\affiliation[e]{Key Laboratory of Quark and Lepton Physics (MOE) and Institute of Particle Physics, Central China Normal University, Wuhan 430079, China}

\emailAdd{hfujii@phys.c.u-tokyo.ac.jp}
\emailAdd{cyrille.marquet@polytechnique.edu}
\emailAdd{watanabe@jlab.org}

\date{\today}

\abstract{
For studying small-$x$ gluon saturation in forward dijet production in high-energy dilute-dense collisions, the improved TMD (ITMD) factorization formula was recently proposed.
In the Color Glass Condensate (CGC) framework, it represents the leading term of an expansion in inverse powers of the hard scale. It contains the leading-twist TMD factorization formula relevant for small gluon's transverse momentum $k_t$, but also incorporates an all-order resummation of kinematical twists, resulting in a proper matching to high-energy factorization at large $k_t$.
In this paper, we evaluate the accuracy of the ITMD formula quantitatively, for the case of quark dijet production in high-energy proton-proton($p+p$) and proton-nucleus ($p+A$) collisions at LHC energies. We do so by comparing the quark-antiquark azimuthal angle $\Delta\phi$ distribution to that obtained with the CGC formula.
For a dijet with each quark momentum $p_t$ much larger than the target saturation scale, $Q_s$, the ITMD formula is a good approximation to the CGC formula in a wide range of azimuthal angle. It becomes less accurate as the jet $p_t$'s are lowered, as expected, due to the presence of genuine higher-twists contributions in the CGC framework, which represent multi-body scattering effects absent in the ITMD formula.
We find that, as the hard jet momenta are lowered, the accuracy of ITMD start by deteriorating at small angles, in the high-energy-factorization regime, while in the TMD regime near $\Delta\phi=\pi$, very low values of $p_t$ are needed to see differences between the CGC and the ITMD formula. In addition, the genuine twists corrections to ITMD become visible for higher values of $p_t$ in $p+A$ collisions, compared to $p+p$ collisions, signaling that they are enhanced by the target saturation scale.
}

\begin{document} 
\maketitle
\flushbottom

\section{Introduction}{\label{sec:Introduction}}

Parton saturation at small Bjorken's $x$ in hadron wave functions 
is one of the most salient and universal features of QCD dynamics\,\cite{Gribov:1984tu,Mueller:1985wy,Mueller:2001fv}.
Small-$x$ partons are interpreted as short-lived quantum fluctuations
splitting from larger-$x$ partons in a hadron wave function.
Lorentz time-dilation dictates that the higher the collision energy is, the smaller-$x$ partons come to
participate in the interaction. The $x$-evolution of the gluon density has been formulated as
the Jalilian-Marian-Iancu-McLerran-Weigert-Leonidov-Kovner (JIMWLK) equation\,
\cite{JalilianMarian:1997jx,JalilianMarian:1997dw,Iancu:2000hn,Ferreiro:2001qy,Weigert:2000gi,Kovner:2013ona},
or the Balitsky-Kovchegov (BK) equation\,\cite{Balitsky:1995ub,Kovchegov:1999yj} in a mean-field approximation.
The evolution changes from a linear to a non-linear character when the gluon density
becomes so dense that the gluon merging starts to compete with the splitting.
This transition is characterized by the so-called saturation momentum scale, 
$Q_s(x)$\,\cite{Gribov:1984tu,Mueller:1985wy,Mueller:2001fv}, an emergent scale in QCD dynamics.
Then, the color-glass-condensate (CGC) framework\,\cite{Iancu:2003xm,Gelis:2010nm,Kovchegov:2012mbw,Albacete:2014fwa},
which describes the small-$x$ part of the wave function in the
presence of large-$x$ random color sources,
has been realized as a suitable effective theory
to calculate observables in the dense gluon regime
with $Q_s(x) \gg \Lambda_{\mathrm{QCD}}$.

Forward dijet production in proton-proton ($p+p$) and proton-nucleus
($p+A$) collisions at the large hadron collider
(LHC) is a unique and valuable observable among others for the phenomenological study of gluon saturation.
In this process a large-$x$ parton from the projectile, which is dilute and well understood in perturbative QCD,
probes the small-$x$ partons in the dense target, and then produces jets at forward rapidities.
This setup is sometimes called \emph{dilute-dense} system.
In addition to its ever highest collision energy,
the nuclear target option available at the LHC
is very advantageous since gluon saturation, or its scale $Q_s(x)$,
is enhanced by the target thickness $\propto A^{1/3}$ ($A$ is the
nuclear mass number). 

In the CGC framework, the dijet production cross-section is expressed in terms of the Wilson line correlators averaged
over external color source distributions. The Wilson-line correlators with fixed transverse positions are essential components
to define the gauge-invariant matrix elements. Those correlators encode multiple scatterings of the partons traversing the dense
target and satisfy the BK-JIMWLK evolution, provided that leading logarithms in $x$ are predominant over leading logarithms in $Q^2$. 
Those multiple scattering effects are enhanced in the dense regime where the saturation scale $Q_s$ increases.
It is demonstrated in Ref.\,\cite{Kotko:2015ura,vanHameren:2016ftb}
that the description of dijet production at the LHC
should simplify thanks to the hard scales involved there.

Indeed, the dijet production contains three characteristic momentum scales:
the typical transverse momentum of a hard jet ${\bm P_t}$, the transverse
momentum imbalance of the pair ${\bm k_t}$,
and the saturation scale of the target $Q_s$.
Here ${\bm P_t}$ is always the hardest scale, while $Q_s$ is the softest of the three.
The original CGC framework does not assume any ordering in the three momentum scales.
In the $Q_s \ll \absol{{\bm P_t}}  \sim \absol{{\bm k_t}}$ limit,
expanding the Wilson line correlators
in the CGC expression to the second order in the gluon field, 
one can obtain the "dilute" result known as high-energy factorization (HEF) or $k_t$-factorization.
On the other hand, in the $Q_s \sim \absol{{\bm k_t}} \ll \absol{{\bm P_t}}$ limit,
by keeping the leading $1/\absol{{\bm P_t}}$ terms from the CGC expression,
one can accurately reproduce the leading-twist TMD factorization result at small $x$ which comes with on-shell hard matrix elements.

In the meantime, by introducing the off-shell $k_t$ dependence of the small-$x$ gluons in the hard matrix elements,
Ref.\,\cite{Kotko:2015ura,vanHameren:2016ftb} proposed an improved TMD (ITMD) expression, which
is valid for any $\absol{{\bm k_t}}$ provided $Q_s \ll \absol{{\bm P_t}}$, and interpolates the TMD and HEF expressions.
Then it was pointed out in Refs.\,\cite{Altinoluk:2019fui,Altinoluk:2019wyu} that such off-shell effect results from the resummation of power
corrections in $\absol{{\bm k_t}}/\absol{{\bm P_t}}$ in the hard scattering parts, known as \emph{kinematic-twists} corrections,
coupled to leading-twist TMD distributions. Alternatively, the ITMD framework can also be thought of as an improvement of HEF, from that
perspective the HEF framework gets supplemented with leading-twist saturation corrections. The ITMD framework provides a concise and useful approximation to the
CGC expression for $Q_s \ll \absol{{\bm P_t}}$, and it is crucial now to assess the quantitative accuracy of the ITMD
formula, compared to the ``full'' CGC formula, when calculating the spectrum of forward dijets.
This is a practical motivation of this paper.

Gluon saturation affects particle production in hadron collisions through the non-linear evolution of the gluon density, 
and through the multiple scattering of the partons with the dense target.
The multiple scattering effects are further categorized into two classes:
the leading-twist ones accounted for in the (I)TMD framework, controlled by the magnitude of $\absol{{\bm k_t}}$ vs. $Q_s$, and those due to genuine higher-twist effects, controlled by $\absol{{\bm P_t}}$ vs. $Q_s$.
The CGC formula contains both effects of multiple scatterings, while the ITMD formula is obtained from the CGC one by getting rid of the genuine higher-twist corrections, which may be referred to as Wandzura-Wilczeck approximation\,\cite{Wandzura:1977qf}. 
The numerical comparison of the ITMD to the CGC formula
will give valuable information about the genuine higher-twist effects
on forward dijets production in high-energy $p+A$ collisions.   
In order to make our ITMD/CGC comparison feasible and clear, we shall restrict our analysis to the forward quark ($q\bar q$) dijet production, and work within the Gaussian truncation of JIMWLK evolution and large-$N_c$ limit, for which the CGC expression is less complicated and can be evaluated directly (indeed, as we will see below, the two expressions then differ only in their hard factors). In this regard, we note that genuine-twist corrections were also analyzed recently in the context of dijet production in deep-inelastic scattering\,\cite{Mantysaari:2019hkq}, using the same approximation but keeping finite $N_c$ corrections.

The paper is organized as follows;
Section\,\ref{sec:Frameworks} gives an overview of the ITMD
and CGC frameworks for forward dijet production. In
Section\,\ref{sec:Results}, we present numerical results on the dijet azimuthal
angle correlation in the ITMD and CGC frameworks. In
particular, we will look into the dependence of the genuine-twist corrections
on kinematics and system size there.
Section\,\ref{sec:Summary} is devoted to summary and concluding remarks.

\section{Frameworks}{\label{sec:Frameworks}}

This section runs through some details of the ITMD and CGC frameworks for forward dijet production in dilute-dense collisions.

\subsection{Improved TMD factorization for forward dijet production}{\label{subsec:ITMD}}

\begin{figure}[t]
\centering
\includegraphics[width=\textwidth]{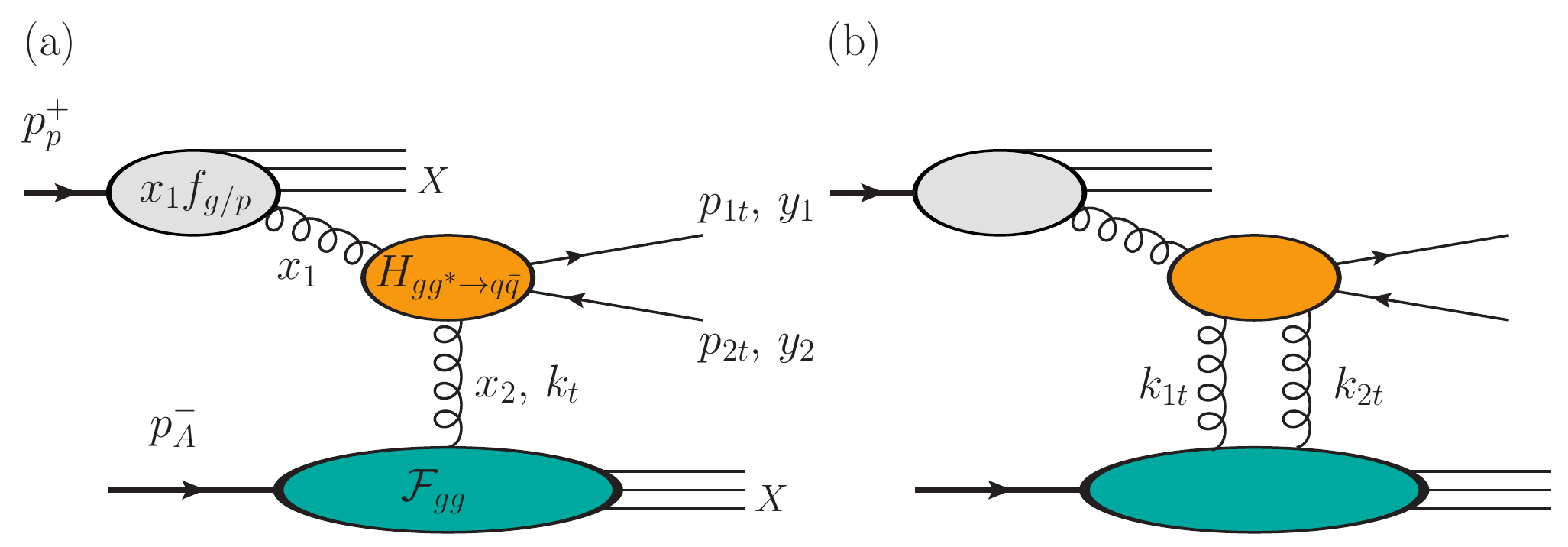}
\caption{
Amplitude-level diagrams for forward quark dijet production $p (p_p) + A (p_A) \to q (p_1) + \bar{q} (p_2)+ X$ from the point of view of the ITMD framework. 
(a): squaring the amplitude provides the $q\bar{q}$ dijet production cross section in the ITMD framework.
(b): diagram yielding so-called genuine-twists corrections, ${\cal O}(Q_s/\absol{{\bm P_t}})$, neglected in the ITMD formula but included in the CGC framework.
}
\label{fig:diagrams}
\end{figure}

We consider the process of inclusive dijet production at forward rapidity in proton-nucleus collisions
\begin{align}
p (p_p) + A (p_A) \to j_1 (p_1) + j_2 (p_2)+ X\ ,
\end{align}
where the four-momenta of the projectile and the target are massless and purely longitudinal. In terms of the light cone variables, $x^\pm = (x^0\pm x^3)/\sqrt{2}$, 
they take the simple form $p_p^\mu\!=\!\sqrt{s/2}\, (1,0,{\bm 0_t})$ and $p_A^\mu\!=\!\sqrt{s/2}\, (0,1,{\bm 0_t})$ where $s$ is the squared center of mass energy (per nucleon-nucleon collisions) of the $p+A$ system. 
The longitudinal momentum fractions of the incoming parton from the projectile, $x_1$, and of the gluon from the target, $x_2$, can be expressed in
terms of the rapidities $(y_1,y_2)$ and transverse momenta $({\bm p_{1t}},{\bm p_{2t}})$ of the produced jets as
\begin{align}
x_1  = \frac{p_1^+ + p_2^+}{p_p^+}   = \frac{1}{\sqrt{s}} \left(\absol{{\bm p_{1t}}} e^{y_1}+\absol{{\bm p_{2t}}} e^{y_2}\right),\quad
x_2  = \frac{p_1^- + p_2^-}{p_A^-}   = \frac{1}{\sqrt{s}} \left(\absol{{\bm p_{1t}}} e^{-y_1}+\absol{{\bm p_{2t}}} e^{-y_2}\right)\,.
\label{eq:x1x2}
\end{align}
By looking at jets produced in the forward direction, we effectively select those fractions to be $x_1 \sim 1$ and $x_2 \ll 1$. Since the target $A$ is probed at low $x_2$, the dominant contributions come from the subprocesses in which the incoming parton on the target side is a gluon, meaning there are three possible channels: $qg\to qg$, $gg\to q\bar q$, and $gg\to gg$. Figure\,\ref{fig:diagrams} shows the kinematics for the $gg\to q\bar q$ subprocess in $p+A$ collisions.

The asymmetry of the problem, $x_1\sim 1$ and $x_2\ll 1$, also implies that gluons from the target have a much bigger average transverse momentum (of the order of $Q_s(x_2)$) compared to that of the partons from the projectile (which is of the order of $\Lambda_\mathrm{QCD}$). Therefore we shall always neglect the transverse momentum of the high-$x_1$ partons from the projectile compared to that of the low-$x_2$ gluons from the target. As a result, the parton content of the projectile hadron is described by regular collinear parton distributions $f_{a/p}(x_1,\mu^2)$ (where $\mu$ is the factorization scale) and TMDs are involved only on the target side, with the transverse momentum of those small-$x_2$ gluons being equal to the transverse momentum of jet pair ${\bm k_t}$:
\begin{align}
{\bm k_t} = {\bm p_{1t}}+{\bm p_{2t}}\, .
\label{eq:ktglue}
\end{align}
This simplification is needed to apply the TMD factorization for the dijet process, since for this final state, there is no such factorization with TMDs for both incoming hadrons\,\cite{Collins:2007nk,Rogers:2010dm}.

The ITMD factorization formula reads \,\cite{Kotko:2015ura}
\begin{align}
\frac{d\sigma({p+A\to j_1+j_2+X})}{dy_1 dy_2 d^2{\bm p_{1t}} d^2{\bm p_{2t}}}=\frac{\alpha_{s}^{2}}{(x_1 x_2 s)^{2}}
\sum_{a,c,d} \frac{x_1 f_{a/p}(x_1,\mu^2)}{1+\delta_{cd}}\sum_i H_{ag^*\to cd}^{(i)}({\bm P_t},{\bm k_t})\mathcal{F}_{ag}^{(i)}(x_2,{\bm k_t}) \ ,
\label{eq:itmd}
\end{align}
where several gluon TMDs $\mathcal{F}_{ag}^{(i)}$ are involved, with different operator definitions, i.e. gauge link structures, and each is accompanied by a different hard factor $H_{ag^*\to cd}^{(i)}$. Its validity domain is $Q_s(x_2)\ll \absol{{\bm P_t}}$, where ${\bm P_t}$ is the hard scale of the process, related to the individual jet momenta:
\begin{align}
{\bm P_t}= \frac{p_2^+ {\bm p_{1t}} - p_1^+ {\bm p_{2t}}}{p_1^+ + p_2^+}=(1-z){\bm p_{1t}}-z{\bm p_{2t}}\,,
\label{eq:Pt-def}
\end{align}
with $z=p_1^+/(p_1^+ + p_2^+)$ the longitudinal momentum fraction carried by the jet $j_1$. The improvement with respect to the TMD factorization formula derived in Ref.\,\cite{Dominguez:2011wm} (in the large-$N_c$ limit) and in Ref.\,\cite{Marquet:2016cgx} (keeping $N_c$ finite), lies in the fact that the hard factors $ H_{ag^*\to cd}^{(i)}({\bm P_t},{\bm k_t})$ are $k_t$-dependent, as opposed to a function of ${\bm P_t}$ only in the TMD case: $H_{ag\to cd}^{(i)}({\bm P_t})=H_{ag^*\to cd}^{(i)}({\bm P_t},{\bm 0_t})$; their expressions can be found in Ref.\,\cite{Kotko:2015ura}. On the other hand, the improvement with respect to the HEF lies in the fact that several gluon distributions are involved, which differ from one another when non-linear effects become important. The various operator definitions of the gluon TMDs $\mathcal{F}_{ag}^{(i)}(x_2,{\bm k_t})$ are found in Ref.\,\cite{Marquet:2016cgx}.

From now on, we focus solely on a quark dijet pair ($q\bar{q}$) production, since considering this subprocess will allow us to make a detailed comparison with the CGC formulation. In that case, let us write down more explicitly the ITMD formula \footnote{Compared to Ref.\,\cite{Marquet:2016cgx}, ${\mathcal F}_{gg}^{(1)}$ is simply denoted ${\cal F}_{gg}$, the Weizs\"{a}cker-Williams gluon TMD ${\mathcal F}_{gg}^{(3)}$ is denoted ${\mathcal F}_{WW}$, and ${\mathcal F}_{\rm adj}={\mathcal F}_{gg}^{(1)}-{\mathcal F}_{gg}^{(2)}$ is the adjoint-dipole gluon TMD\,\cite{Marquet:2017xwy}.};
\begin{align}
\frac{d\sigma(pA\to q\bar{q}X)}{dy_1 dy_2 d^2{\bm p_{1t}} d^2{\bm p_{2t}}} 
= \frac{\alpha_s^2}{2C_F}\frac{z(1-z)}{p_{1t}^2 p_{2t}^2}
x_1f_{g/p}(x_1,\mu^2)&P_{qg}(z)\left[{\mathcal F}_{gg}(x_2,{\bm k_{t}})-\frac1{N_c^2}{\mathcal F}_{WW}(x_2,{\bm k_{t}})\right.\non
&\left.
+\frac{2z(1-z){\bm p_{1t}}\cdot {\bm p_{2t}}}{P_t^2}{\mathcal F}_{\rm adj}(x_2,{\bm k_{t}})\right]
\, ,
\label{eq:itmd-qqbar}
\end{align}
where 
\begin{align}
P_{qg}(z) = \frac{z^2+(1-z)^2}{2}
\end{align}
denotes the usual gluon-quark splitting function at leading order in $\alpha_s$.
The relevant small-$x$ gluon TMDs are given by \,\cite{Marquet:2016cgx}
\begin{align}
\mathcal{F}_{gg}(x_2,{\bm k_{t}}) &= \frac{4}{g^2}\int\frac{d^2{\bm x} d^2{\bm y}}{(2\pi)^3}\ e^{-i{\bm k_t}\cdot({\bm x}-{\bm y})}
\frac1{N_c}
\left\langle \mathrm{Tr}\left[  (\partial_i U_{\bm y}) (\partial_i U^\dagger_{\bm x}) \right] \mathrm{Tr}\left[ U_{\bm x} U^\dagger_{\bm y}\right] \right\rangle_{x_2}\ ,
\nonumber\\
\mathcal{F}_{\rm adj}(x_2,{\bm k_{t}}) &= \frac{2}{g^2} \int\frac{d^2{\bm x} d^2{\bm y}}{(2\pi)^3}\ e^{-i{\bm k_t}\cdot({\bm x}-{\bm y})}
\frac1{N_c}
\left\langle \mathrm{Tr}\left[  (\partial_i V_{\bm y}) (\partial_i V^\dagger_{\bm x}) \right] \right\rangle_{x_2}\ ,
\nonumber\\
\mathcal{F}_{WW}(x_2,{\bm k_{t}}) &= -\frac{4}{g^2}\int\frac{d^2{\bm x} d^2{\bm y}}{(2\pi)^3}\ e^{-i{\bm k_t}\cdot({\bm x}-{\bm y})}
\left\langle \mathrm{Tr}\left[ (\partial_i U_{\bm x}) U^\dagger_{\bm y} (\partial_i U_{\bm y}) U^\dagger_{\bm x}  \right] \right\rangle_{x_2}\ ,
\label{eq:gluontmds}
\end{align}
in terms of the Wilson lines
\begin{equation}
U_{{\bm x}}=\mathcal{P}\exp\left[ig_s\int_{-\infty}^{\infty}\mathrm{d}x^{+}A_{a}^{-}(x^{+},{\bm x})t^{a}\right]\;,\quad V_{{\bm x}}=\mathcal{P}\exp\left[ig_s\int_{-\infty}^{\infty}\mathrm{d}x^{+}A_{a}^{-}(x^{+},{\bm x})T^{a}\right]
\end{equation}
with $t^{a}$ and $T^{a}$ denoting the generators of the fundamental and adjoint representation of $SU(N_{c})$, respectively. The process is depicted in Fig.\,\ref{fig:diagrams} at the amplitude level, the ITMD cross-section being the square of Fig.\,\ref{fig:diagrams} (a). The soft gluons attaching to the hard parts, are not shown, those are accounted for by two (fundamental) Wilson lines. A derivative applied to a Wilson line corresponds to a gluon exchanged in the $t$-channel, those are explicitly drawn.

%%%%%%%%%%%%%%%%%%%%%%% figure %%%%%%%%%%%%%%%%%%%%
%%%%%%%%%%%%%%%%%%%%%%% figure %%%%%%%%%%%%%%%%%%%%

The CGC averages $\langle\ \cdot\ \rangle_{x_2}$ represent averages over the configurations of the classical color field of the hadronic/nuclear target, $A^-$, which describes the dense parton content of its wave function, at small longitudinal momentum fraction $x_2$. In the leading-logarithmic approximation, the evolution of the CGC averages with decreasing $x_2$ obeys the JIMWLK equation,
\begin{align}
\frac{d}{d\ln(1/x_2)} \left\langle O \right\rangle_{x_2}  = \left\langle H_{\text{JIMWLK}}\ O \right\rangle_{x_2}
\end{align} 
where $H_{\text{JIMWLK}}$ denotes the JIMWLK Hamiltonian.

The ITMD formula \eqref{eq:itmd} is an interpolation between two limiting cases, $Q_s\ll \absol{{\bm k_t}},\absol{{\bm P_t}}$ and $Q_s, \absol{{\bm k_t}} \ll \absol{{\bm P_t}}$, both limits being contained as well in the more general CGC framework (the details of which are recalled below). The ITMD formula is valid when $\absol{{\bm P_t}}\gg Q_s(x_2)$, however the value of $\absol{{\bm k_t}}$ can be arbitrary. When $\absol{{\bm k_t}}\gg Q_s(x_2)$, the HEF formula (aka $k_t$-factorization) is recovered: the various gluon TMDs \eqref{eq:gluontmds} collapse into a single function, known as the unintegrated gluon distribution, which evolves according to the Balitsky-Fadin-Kuraev-Lipatov (BFKL) evolution equation\,\cite{Lipatov:1976zz,Kuraev:1976ge,Balitsky:1978ic}. By contrast, the TMD factorization formula emerges from \eqref{eq:itmd} when $\absol{{\bm k_t}}\sim Q_s(x_2)$, it is formally obtained by replacing $H_{ag^*\to cd}^{(i)}({\bm P_t},{\bm k_t})$ with $H_{ag^*\to cd}^{(i)}({\bm P_t},{\bm 0_t})$; in that regime, (leading-twist) non-linear effects are important, and induce significant differences between the gluon TMDs.

Starting from the TMD formula, restoring the off-shellness of the small-$x$ gluons in the hard factors and hereby obtaining the ITMD formula is equivalent to performing an all-order resummation of power corrections in $ \absol{{\bm k_t}} / \absol{{\bm P_t}}$, known as kinematical-twists corrections\,\cite{Altinoluk:2019fui}. Furthermore, the difference between the ITMD formula and the more complete CGC formulation represents corrections of the genuine-twists kind\,\cite{Altinoluk:2019wyu}, that should become important when $\absol{{\bm P_t}}\sim Q_s(x_2)$. Diagrammatically, those genuine-twist corrections come from Fig.\,\ref{fig:diagrams} (b), meaning 3-body and 4-body terms after squaring. At the cross-section level, all contributions in Fig.\,\ref{fig:diagrams} involve 4 Wilson lines (4 fundamental ones in the case of the $q\bar q$ final state considered here), but the 3- (resp. 4-) body contribution involves 3 (resp. 4) derivatives and 3 (resp. 4) different transverse positions, while the ITMD cross-section is a two-body contribution which involve 2 derivatives and 2 different transverse positions, as is explicit in \eqref{eq:gluontmds}.

\subsection{CGC framework for forward $q\bar q$ pair production}{\label{subsec:CGC}}

\begin{figure}[t]
\centering
\includegraphics[width=\textwidth]{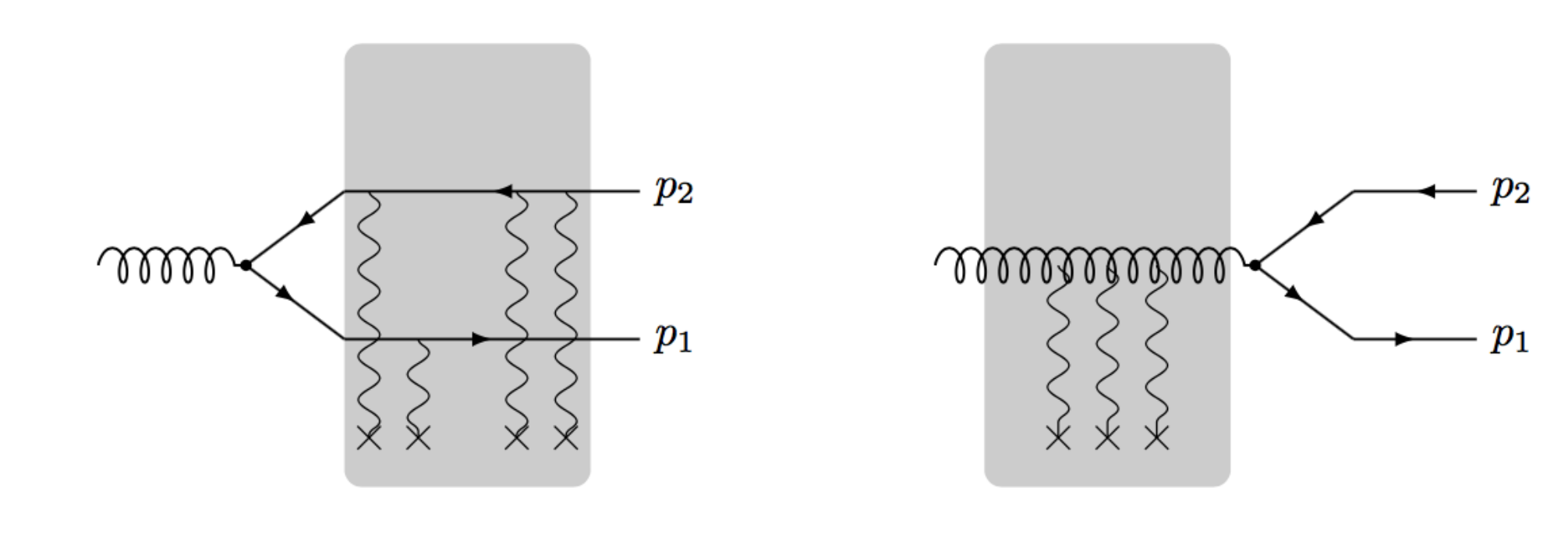}
\caption{
Quark-pair production amplitude in the CGC formalism, in which the pair is radiated from the gluon before (left) or after (right) the multiple interactions with the gauge fields in the target represented by wavy lines. Each propagating parton picks up a Wilson line, implying 2-, 3-, and 4-point Wilson line correlators after squaring.
}
\label{fig:cgcdiagrams}
\end{figure}

In this subsection, we recall the CGC formalism for $q\bar q$ pair production in dilute-dense collisions. In the amplitude and complex conjugate amplitude, the incoming gluon from the dilute projectile may split into the $q\bar q$ pair before or after the interaction with the dense target, as pictured in Fig.\,\ref{fig:cgcdiagrams}. Fundamental Wilson lines describe the interaction for quarks, and adjoint Wilson lines for gluons. As a result, the cross-section involves four contributions: a correlator of four fundamental Wilson lines, $S^{(4)}$, corresponding to interactions happening after the gluon splitting into the $q\bar q$ pair, both in the amplitude and the complex conjugate amplitude; a correlator of two adjoint Wilson lines, $S^{(2)}$, corresponding to interactions taking place before the gluon splitting, both in the amplitude and the complex conjugate amplitude; two correlators of three Wilson lines, $S^{(3)}$, for the interference terms.

%%%%%%%%%%%%%%%%%%%%%%% figure %%%%%%%%%%%%%%%%%%%%
%%%%%%%%%%%%%%%%%%%%%%% figure %%%%%%%%%%%%%%%%%%%%

Denoting $p$ the momentum of the incoming gluon, the cross-section reads \,\cite{Dominguez:2011wm}:
\begin{align}
&\frac{d\sigma(pA\to q\bar{q} X)}{dy_1 dy_2 d^2{\bm p_{1t}} d^2{\bm p_{2t}}} = \frac{\alpha_s}{2}z(1-z)x_1 f_{g/p}(x_1,\mu^2)
\int\frac{d^2{\bm u}}{(2\pi)^2}\frac{d^2{\bm u'}}{(2\pi)^2} e^{i {\bm P_t} \cdot ({\bm u'}-{\bm u})}\non
\times&p^+\!\sum_{\lambda\alpha\beta} \varphi^{\lambda^*}_{\alpha\beta}(p,p_1^+,{\bm u'}) \varphi^{\lambda}_{\alpha\beta}(p,p_1^+,{\bm u}) 
\int\frac{d^2{\bm v}}{(2\pi)^2}\frac{d^2{\bm v'}}{(2\pi)^2} e^{i {\bm k_t} \cdot ({\bm v'}-{\bm v})}
\left\{S^{(4)}_{q\bar{q}\bar{q}q}\left({\bm x},{\bm b},{\bm x'},{\bm b'};x_2\right)\right.\non
&\left.-S^{(3)}_{qg\bar{q}}\left({\bm x},{\bm v'},{\bm b};x_2\right)
-S^{(3)}_{qg\bar{q}}\left({\bm b'},{\bm v},{\bm x'},x_2\right)+S^{(2)}_{gg}\left({\bm v},{\bm v'};x_2\right)\right\}\ ,
\label{eq:cgc-qqbar}
\end{align}
where
\begin{align}
{\bm x}={\bm v}+(1\!-\!z){\bm u}\quad\mbox{and}\quad{\bm x'}={\bm v'}+(1\!-\!z){\bm u'}
\end{align}
denote the transverse positions of the final-state quark in the amplitude and the conjugate amplitude, respectively, and
\begin{align}
{\bm b}={\bm v}-z{\bm u}\quad\mbox{and}\quad{\bm b'}={\bm v'}-z{\bm u'}
\end{align}
denote the transverse positions of the final-state antiquark in the amplitude and the conjugate amplitude, respectively. 
${\bm u'}-{\bm u}$ is conjugate to the hard momentum ${\bm P_t}=(1\!-\!z){\bm p_{1t}}-z{\bm p_{2t}}$, and ${\bm v'}-{\bm v}$ is conjugate to the total transverse momentum of the produced particles ${\bm k_t}={\bm p_{1t}}+{\bm p_{2t}}$.

The $S^{(i)}$ Wilson line correlators are given by
\begin{align}
S^{(4)}_{q\bar{q}\bar{q}q}({\bm x},{\bm b},{\bm x'},{\bm b'};x_2)
&=\frac{1}{C_F N_c}\left<{\text {Tr}} \left(U^\dagger_{\bm b} t^c U^{\phantom \dagger}_{\bm x} U^\dagger_{\bm x'} t^c U^{\phantom \dagger}_{\bm b'}\right)\right>_{x_2}\ ,\\
S^{(3)}_{qg\bar{q}}({\bm x},{\bm v},{\bm b};x_2)&=\frac{1}{C_F N_c}\left<\text {Tr}\left(U^\dagger_{\bm b}t^cU^{\phantom \dagger}_{\bm x}t^d\right)V^{cd}_{\bm v}\right>_{x_2}\ ,\\
S^{(2)}_{gg}({\bm v},{\bm v'};x_2)&=\frac{1}{N^2_c-1}\left<{\text {Tr}} \left(V^{\phantom \dagger}_{\bm v} V^\dagger_{\bm v'} \right)\right>_{x_2}\ .
\end{align}
The functions $\varphi^{\lambda}_{\alpha\beta}$ denote the $g\to q\bar q$ splitting wave functions. 
In the limit of massless quarks, the wave function overlap is simply given by
\begin{align}
p^+ \sum_{\lambda\alpha\beta} \varphi^{\lambda^*}_{\alpha\beta}(p,p^+_1,{\bm u'}) \varphi^{\lambda}_{\alpha\beta}(p,p^+_1,{\bm u}) =
16\pi^2 \frac{{\bm u} \cdot {\bm u'}}{\absol{\bm u}^2  \absol{\bm u'}^2} P_{qg}(z) \ .
\end{align}
The three scales $Q_s$, $\absol{\bm k_t}$, and $\absol{\bm P_t}$ are characterizing the kinematics for the dijet production. It is instructive to consider the two limits $Q_s\ll \absol{\bm k_t}, \absol{\bm P_t}$ and $Q_s, \absol{\bm k_t}\ll \absol{\bm P_t}$ in the CGC framework.

It was shown in Ref.\,\cite{Kotko:2015ura} that in the $Q_s\ll \absol{\bm k_t}\sim \absol{\bm P_t}$ limit, the formula \eqref{eq:cgc-qqbar} reduces to
\begin{align}
p_{1t}^2\ p_{2t}^2\frac{d\sigma(pA\to q\bar{q}X)}{dy_1 dy_2 d^2{\bm p_{1t}} d^2{\bm p_{2t}}} =&
\frac{\alpha_s^2}{2C_F}x_1f_{g/p}(x_1,\mu^2)z(1\!-\!z)P_{qg}(z)\non
\times&\left[\frac{(1\!-\!z)^2 p_{1t}^{\ 2}+z^2 p_{2t}^{\ 2}}{P_t^2}-\frac{1}{N_c^2}\right]
{\mathcal F}^{\rm dilute}_{g/A} (x_2,{\bm k_t})
\label{eq:sigmaCGCfac-qbarq}
\end{align}
where
\begin{align}
{\mathcal F}^{\rm dilute}_{g/A} (x_2,{\bm k_t})= 4 \int\frac{d^3xd^3y}{(2\pi)^3}\ e^{-i{\bm k_t}\cdot({\bm x}-{\bm y})}
\left\langle {\mathrm {Tr}} [\partial_i A^-(x^+, {\bm x})] [\partial_i (A^-(y^+, {\bm y})]\right\rangle_{x_2}\ .
\end{align}
It corresponds to the BFKL limit of the CGC and is referred to as the HEF formula. It has been extensively studied in the literature\,\cite{Deak:2009xt,Kutak:2012rf,vanHameren:2013fla,vanHameren:2014lna,vanHameren:2014ala} 
(where the gluon TMD is denoted by ${\cal F}_{g/A}=\pi \Phi_{g/A}$ due to a different normalization convention)
\footnote{As a related topic, Ref.\,\cite{Watanabe:2016gws} clarifies how the BFKL evolution equation appears for forward hadron production in the hybrid CGC formula with a dilute target.}.  
Its domain of validity corresponds to jets produced away from the back-to-back region, where the small-$x_2$ gluon is hard, and saturation effects are negligible. 
However, provided that we are dealing with forward jets, linear small-$x$ effects are still relevant\,\cite{vanHameren:2014ala}.

In the meantime, it was shown in Ref.\,\cite{Dominguez:2011wm,Marquet:2016cgx} that in the $Q_s\sim \absol{\bm k_t}\ll \absol{\bm P_t}$ limit, the formula \eqref{eq:cgc-qqbar} becomes
\begin{align}
p_{1t}^2\ p_{2t}^2\frac{d\sigma(pA\to q\bar q X)}{dy_1 dy_2 d^2{\bm p_{1t}} d^2{\bm p_{2t}}} = \frac{\alpha_s^2}{2C_F}z(1\!-\!z)
x_1f_{g/p}(x_1,\mu^2)P_{qg}(z)\biggl[{\mathcal F}_{gg}(x_2,{\bm k_{t}}) \non
\left.-\frac1{N_c^2}{\mathcal F}_{WW}(x_2,{\bm k_{t}})-2z(1\!-\!z){\mathcal F}_{\rm adj}(x_2,{\bm k_{t}})\right]\,.
\label{eq:fullTMD}
\end{align}
This is a TMD factorization, obtained from the CGC by extracting the leading $1/\absol{\bm P_t}$ power. Its domain of validity corresponds to nearly back-to-back jets production. In our small-$x$ context (forward jets), saturation effects in $Q_s/\absol{\bm k_t}$ must be accounted for here, without them the TMDs all coincide. We note that the TMD approach has been previously extensively studied in the literature\,\cite{Bomhof:2006dp,Boer:1999si,Belitsky:2002sm,Boer:2003cm,Collins:2007nk,Vogelsang:2007jk,Rogers:2010dm,Xiao:2010sp} in a broader context than small-$x$ physics, in which case the process dependence of the TMDs is simply non-perturbative ($Q_s$ is not large enough compared to $\Lambda_\textrm{QCD}$).

The ITMD factorization formula \eqref{eq:itmd} (as well as those for the other two channels) was built in order to contain both those expressions as its limiting cases,
as therefore be valid regardless of the magnitude of $\absol{\bm k_t}$. In the first case, it is so because in the $Q_s\ll \absol{\bm k_t}$ limit, one has $\mathcal{F}_{gg},\mathcal{F}_{\rm adj},\mathcal{F}_{WW}\to {\mathcal F}^{\rm dilute}_{g/A} (x_2,{\bm k_t})+{\mathcal O}(1/k_t^2)$. In the second case, it occurs because in the $\absol{\bm k_t} \ll \absol{\bm P_t}$ limit the coefficient in front of $\mathcal{F}_{\rm adj}$ becomes $-2z(1-z)$. We note that, any systematic improvements of the HEF or TMD factorization frameworks in perturbation theory, which may be obtained in the future, could be implemented in the ITMD factorization formula as well.

Finally, the difference between the CGC formula \eqref{eq:cgc-qqbar} and the ITMD formula \eqref{eq:itmd-qqbar} was clarified recently \,\cite{Altinoluk:2019fui}. The CGC amplitude pictured in Fig.\,\ref{fig:cgcdiagrams}, whose square leads to \eqref{eq:cgc-qqbar}, can be rewritten in an alternative way using an expansion in the dipole sizes conjugate to ${\bm P_t}$, which corresponds to a twist expansion. The leading contribution represented in Fig.\,\ref{fig:diagrams} (a) (whose square leads to the ITMD formula \eqref{eq:itmd-qqbar}), is made of the leading $1/\absol{\bm P_t}$ term (the TMD term extracted in \cite{Dominguez:2011wm,Marquet:2016cgx}) and an all-order resummation of a sub-set of higher-order terms, the so-called kinematical twists of order ${\mathcal O}(\absol{\bm k_t}/\absol{\bm P_t})$ (whose effect at the cross-section level is to restore the off-shellness of the gluon in the hard factor while leaving the leading-twist TMD structure unchanged). The remaining higher-order contributions represented in Fig.\,\ref{fig:diagrams} (b), of order ${\mathcal O}(Q_s/\absol{\bm P_t})$, represent the difference between the CGC and the ITMD formula, they are known as genuine twists terms. Our goal now is to estimate the magnitude of that difference. To do that, we shall consider the large-$N_c$ limit, in which case the CGC framework becomes tractable, especially with the $gg \to q\bar{q}$ channel.

\subsection{ITMD/CGC comparison in the large-$N_c$ limit}{\label{subsec:Large-Nc}}

To enable an ITMD/CGC comparison easily, let us simplify Eqs.\,\eqref{eq:itmd-qqbar} and \eqref{eq:cgc-qqbar}. As for the multi-point correlators in \eqref{eq:cgc-qqbar}, in terms of the fundamental Wilson lines, we can write down those as
\begin{align}
S^{(4)}_{q\bar{q}\bar{q}q}({\bm x},{\bm b},{\bm x'},{\bm b'};x_2)&=
\frac{N_c}{2C_F}\left < D({\bm x},{\bm x'})D({\bm b'},{\bm b}) -\frac{1}{N_c^2} Q({\bm x},{\bm x'},{\bm b'},{\bm b}) \right>_{x_2}\ ,\label{eq:four-point}\\
S^{(3)}_{qg\bar{q}} ({\bm x},{\bm v},{\bm b};x_2) &= \frac{N_c}{2C_F} \left < D({\bm x},{\bm v})D({\bm v},{\bm b}) -\frac{1}{N_c^2}D({\bm x},{\bm b})\right>_{x_2}\ ,\label{eq:three-point}\\
S^{(2)}_{gg}({\bm v},{\bm v'};x_2)&=\frac{N_c}{2C_F}\left < D({\bm v},{\bm v'})D({\bm v'},{\bm v}) -\frac{1}{N_c^2}\right>_{x_2}\ ,\label{eq:two-point}
\end{align}
where 
\begin{align}
D({\bm x},{\bm y})=\frac{1}{N_c}{\mathrm{Tr}}\left(U_{\bm x} U^\dagger_{\bm y}\right)\quad\mbox{and}\quad
Q({\bm x},{\bm y},{\bm v},{\bm w})=\frac{1}{N_c}{\mathrm {Tr}}\left(U_{\bm x} U^\dagger_{\bm y} U_{\bm v} U^\dagger_{\bm w}\right)\ .
\end{align}
Eqs.\,\eqref{eq:four-point}, \eqref{eq:three-point}, and \eqref{eq:two-point}  are still complicated for a clear comparison of the two approaches. To make the multi-point correlators more manageable, we shall utilize the so-called Gaussian approximation of the CGC\,\cite{Fujii:2006ab,Marquet:2007vb,Kovchegov:2008mk,Marquet:2010cf,Dumitru:2011vk,Iancu:2011nj,Alvioli:2012ba}.  The essential point is to assume that all the color charge correlations in the target stay Gaussian throughout the evolution. This is found to be a reasonable approximation to the multi-point correlators obtained from the JIMWLK evolution\,\cite{Dumitru:2011vk,Alvioli:2012ba}.
On top of the Gaussian approximation, for simplicity, we shall work in the large-$N_c$ limit.
In addition to dropping the explicitly large-$N_c$ suppressed terms,
this allows to write a correlator of a product of traces as the product of single trace correlators. Thus, the combination inside the brackets $\big\{ \cdot \big\}$ in Eq.\,\eqref{eq:cgc-qqbar} can be cast into
\begin{align}
\frac{N_c}{2C_F}\Big\{
S_{q\bar q}[{\bm v}\!+\!(1\!-\!z){\bm u},{\bm v'}\!+\!(1\!-\!z){\bm u'};x_2]S_{q\bar q}[{\bm v'}\!-\!z{\bm u'},{\bm v}\!-\!z{\bm u};x_2]+
S_{q\bar q}[{\bm v},{\bm v'};x_2]S_{q\bar q}[{\bm v'},{\bm v};x_2]\non
-S_{q\bar q}[{\bm v}\!+\!(1\!-\!z){\bm u},{\bm v'};x_2]S_{q\bar q}[{\bm v'},{\bm v}\!-\!z{\bm u};x_2]-
S_{q\bar q}[{\bm v'}\!-\!z{\bm u'},{\bm v};x_2]S_{q\bar q}[{\bm v},{\bm v'}\!+\!(1\!-\!z){\bm u'};x_2]\Big\}\ ,
\end{align}
in terms of only the two-point function (dipole amplitude) $S_{q\bar q}({\bm x},{\bm y};x_2)=\langle D({\bm x},{\bm y})\rangle_{x_2}$.
We will see below that the above treatment for the multi-point correlators helps us capture differences between the ITMD and CGC framework.

Then, introducing the dipole amplitude in the momentum space,
\begin{align} 
F(x_2,{\bm k_t}) = \int \frac{d^2{\bm r}}{(2\pi)^2}\ e^{-i{\bm k_t}\cdot{\bm r}} S_{q\bar q}({\bm b}+{\bm r}/2,{\bm b}-{\bm r}/2;x_2)\ ,
\label{eq:FTofSqq}
\end{align}
and neglecting the ${\bm b}$ dependence of $F$ for simplicity, the second and third lines of Eq.\,\eqref{eq:cgc-qqbar} simplify into
\begin{align}
S_{\perp}\frac{N_c}{2C_F}\int\frac{d^2{\bm q_t}}{(2\pi)^2}F(x_2,{\bm q_t})F(x_2,{\bm q_t}-{\bm k_t})
\left(1-e^{i({\bm q_t}-z{\bm k_t})\cdot{\bf u}}\right)\left(1-e^{-i({\bm q_t}-z{\bm k_t})\cdot{\bm u'}}\right)\ ,
\end{align}
where $S_{\perp}$ represents the transverse area of the target. Finally, with these approximations the cross section for producing a pair of $q$ at $y_1$ with ${\bm p_{1t}}$ and $\bar{q}$ at $y_2$ with ${\bm p_{2t}}$ in the forward rapidity region is given by
\begin{align}
\frac{d\sigma(pA\to q\bar{q} X)}{dy_1 dy_2 d^2{\bm p_{1t}} d^2{\bm p_{2t}}}=\frac{\alpha_s N_c}{2C_F}& \frac{S_\perp}{8\pi^2}z(1-z)x_1f_{g/p}(x_1,\mu^2)
\int d^2{\bm q_t} F(x_2,{\bm q_t})F(x_2,{\bm q_t}-{\bm k_t})\non
&\times p^+\!\sum_{\lambda\alpha\beta}\left|\tilde\varphi^{\lambda}_{\alpha\beta}(p,p_1^+,{\bm P_t})-\tilde\varphi^{\lambda}_{\alpha\beta}(p,p_1^+,{\bm p_{1t}}-{\bm q_t})\right|^2
\end{align}
with $\tilde\varphi^{\lambda}_{\alpha\beta}(p,p_1^+,{\bm P_t})=\int\frac{d^2{\bm u}}{(2\pi)^2}e^{-i {\bm P_t} \cdot{\bm u}}  \varphi^{\lambda}_{\alpha\beta}(p,p_1^+,{\bm u})$.
In the massless quarks limit, this is simply given by
\begin{align}
p^+\!\sum_{\lambda\alpha\beta}\absol{\tilde\varphi^{\lambda}_{\alpha\beta}(p,p_1^+,{\bm P_t})-\tilde\varphi^{\lambda}_{\alpha\beta}(p,p_1^+,{\bm p_{1t}}-{\bm q_t})}^2
=&\,4P_{qg}(z)\absol{\frac{{\bm P_t}}{P_t^2} - \frac{{\bm p_{1t}}-{\bm q_t}}{({\bm p_{1t}}-{\bm q_t})^2}}^2\non
=&\,4P_{qg}(z)\frac{(z{\bm k_t}-{\bm q_t})^2}{P_t^2({\bm p_{1t}}-{\bm q_t})^2}\,,
\label{eq:splitting-kernel}
\end{align}
where we have used the identity: 
\begin{align}
\int d^2{\bm u}\; e^{i{\bm k_t} \cdot {\bm u}}\frac{{\bm u}}{\absol{\bm u}^2}=2\pi i\; \frac{{\bm k_t}}{\absol{\bm k_t}^2}.
\end{align}
Therefore, provided the large-$N_c$ limit, the CGC formula for forward dijet production reads
\begin{align}
\frac{d\sigma(pA \to q\bar q X)}{dy_1dy_2d^2{\bm p_{1t}}d^2{\bm p_{2t}}}\Big|_{\rm CGC}=
&\frac{\alpha_sS_\perp}{2\pi^2}z(1-z)P_{qg}(z)
\frac{x_1f_{g/p}(x_1,\mu^2)}{P_t^2}
\int d^2{\bm q_t} \,F(x_2,{\bm q_t})F(x_2,{\bm k_t}-{\bm q_t})\non
&\times\left[\frac{(1-z)^2({\bm k_t}-{\bm q_t})^2+z^2q_t^2-2z(1-z){\bm q_t}\cdot ({\bm k_t}-{\bm q_t})}{({\bm q_t}-{\bm p_{2t}})^2}\right]\ .
\label{eq:cgc-large-Nc}
\end{align}
We have performed the change of variable ${\bm q_t}\to {\bm k_t}-{\bm q_t}$ and then wrote $(1-z){\bm k_t}-{\bm q_t}=(1-z)({\bm k_t}-{\bm q_t})-z{\bm q_t}$ before squaring (this choice for writing \eqref{eq:cgc-large-Nc} will make for easier comparisons with the ITMD formula). Also, we have put $N_c/(2C_F)\to 1$ in the overall prefactor.

Next, let us examine the ITMD framework by using the same approximations as illustrated above. With our approximations, the ITMD framework for forward dijet production now involves only two gluon TMDs, and from \eqref{eq:gluontmds}, they can be written as:
\begin{align}
\mathcal{F}_{gg}(x_2,{\bm k_t})=&\,\frac{N_c S_\perp}{2\pi^2\alpha_s}\int d^2{\bm q_t}\; q_t^2\ F(x_2,{\bm q_t})\; F(x_2,{\bm k_t}-{\bm q_t})\,,\\
\mathcal{F}_{\rm adj}(x_2,{\bm k_t})=&\,\frac{N_c S_\perp}{4\pi^2\alpha_s}\int d^2{\bm q_t}\; k_t^2\ F(x_2,{\bm q_t})\;F(x_2,{\bm k_t}-{\bm q_t})\,,
\label{eq:gluontmds-large-Nc}
\end{align}
where we have used $\mathrm{Tr}\left[V_{{\bm v}}V_{{\bm v'}}^{\dagger}\right]=N_{c}^{2}|D({\bm v},{\bm v'})|^{2}-1$. The forward dijet cross section in the ITMD framework is then given by
\begin{align}
\frac{d\sigma(pA\to q\bar q X)}{dy_1 dy_2 d^2{\bm p_{1t}} d^2{\bm p_{2t}}}\Big|_{\rm ITMD}=
&\,\frac{\alpha_sS_\perp}{2\pi^2} z(1-z)P_{qg}(z)\frac{x_1f_{g/p}(x_1,\mu^2)}{P_t^2}
\int d^2{\bm q_t}\; F(x_2,{\bm q_t}) F(x_2,{\bm k_t}-{\bm q_t})\non
&\times\left[\frac{(1-z)^2}{p_{2t}^2} q_t^2+\frac{z^2}{p_{1t}^2} q_t^2+\frac{2z(1-z){\bm p_{1t}}\cdot {\bm p_{2t}}}{p_{1t}^2\ p_{2t}^2} \left(\frac{k_t^2}{2}-q_t^2\right)\right]\,\non
=&\,\frac{\alpha_sS_\perp}{2\pi^2} z(1-z)P_{qg}(z)\frac{x_1f_{g/p}(x_1,\mu^2)}{P_t^2}
\int d^2{\bm q_t}\; F(x_2,{\bm q_t}) F(x_2,{\bm k_t}-{\bm q_t})\non
&\times\left[\frac{(1-z)^2}{p_{2t}^2} ({\bm k_t}-{\bm q_t})^2+\frac{z^2}{p_{1t}^2} q_t^2+\frac{2z(1-z){\bm p_{1t}}\cdot {\bm p_{2t}}}{p_{1t}^2\ p_{2t}^2} {\bm q_t}\cdot({\bm k_t}-{\bm q_t})\right]\,.
\label{eq:itmd-large-Nc}
\end{align}
To reach the second line of Eq.\,\eqref{eq:itmd-large-Nc}, we have used the change of variable ${\bm q_t}\to {\bm k_t}-{\bm q_t}$. This can now be compared with Eq.\,\eqref{eq:cgc-large-Nc}.

Let us emphasize the purpose of this paper again. In this section, we have highlighted the difference 
between the ITMD and CGC frameworks analytically, using the Gaussian truncation and the large-$N_c$ limit: 
the hard scattering part in Eq.\,\eqref{eq:itmd-large-Nc} differs from the one in Eq.\,\eqref{eq:cgc-large-Nc}. Our
interest now is to estimate the genuine twist corrections absent in the former but present in the latter. In the following section, we shall further examine that, numerically.

Before, it is worthwhile to give the HEF and TMD limits using the same simplifications, as we shall numerically evaluate them later as well. In the HEF limit where $\absol{\bm k_t} \gg Q_s$, $S_\perp k^2_t F = (2\pi^2\alpha_s/N_c){\cal F}^{\rm dilute}_{g/A}$, and from 
\eqref{eq:sigmaCGCfac-qbarq} we have
\begin{align}
\frac{d\sigma(pA \to q \bar{q} X)}{dy_1 dy_2 d^2 {\bm p_{1t}} d^2 {\bm p_{2t}}}\Big|_{\rm HEF}= 
\frac{\alpha_sS_\perp}{2\pi^2} z(1-z) P_{qg}(z)
\frac{x_1f_{g/p}(x_1,\mu^2)}{P_t^2}
\left[  \frac{(1-z)^2}{p_{2t}^2} + \frac{z^2}{p_{1t}^2} \right]
k^2_t F(x_2,{\bm k_t})\,.
\label{eq:HEF}
\end{align}
Meanwhile, in the TMD limit, from \eqref{eq:fullTMD} one obtains
\begin{align}
\frac{d\sigma(pA \to q \bar{q} X)}{dy_1 dy_2 d^2 {\bm p_{1t}} d^2 {\bm p_{2t}}}\Big|_{\rm TMD}=
&\frac{\alpha_sS_\perp}{2\pi^2} z(1-z) P_{qg}(z)
\frac{x_1f_{g/p}(x_1,\mu^2)}{p_{1t}^2p_{2t}^2}
\non
&\times \int d^2 {\bm q_t}\;F(x_2,{\bm q_t})F(x_2,{\bm k_t}-{\bm q_t}) [q_t^2 -z(1-z) k_t^2 ]\,.
\label{eq:TMD}
\end{align}

\section{Numerical setup and results}{\label{sec:Results}}

In order to exemplify the accuracy of the ITMD framework, we evaluate
the azimuthal angle correlation in forward $q\bar q$  dijet production with
the ITMD formula \eqref{eq:itmd-large-Nc} and with the CGC formula
\eqref{eq:cgc-large-Nc} in $p+p$ and $p+A$ collisions and compare
these results.

\begin{figure}[t]
\centering
\includegraphics[width=0.6\textwidth]{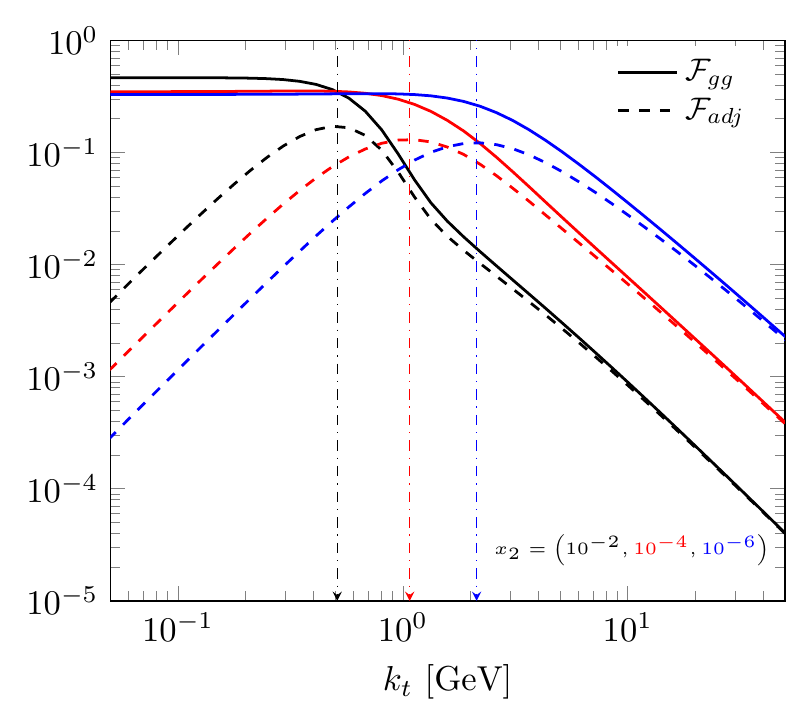}
\caption{Gluon TMDs $\mathcal{F}_{gg}$ (solid lines) and
$\mathcal{F}_{\rm adj}$ (dashed lines) as a function of transverse momentum $k_t=\absol{\bm k_t}$ with fixed $x=10^{-2}$ (black), $10^{-4}$ (red) and $10^{-6}$ (blue).
The pre-factor $\alpha_s/S_\perp$ is omitted.
The saturation scale $Q_s(x)$ is defined here as 
the peak position of $\mathcal{F}_{\rm adj}$,
indicated by a dash-dotted arrow for each $x$.
}
\label{fig:gtmds}
\end{figure}

%%%%%%%%%%%%%%%%%%%%%%%%%%%%%%%%%%%%%%%%%%%%%%%%
\subsection{Setup}{\label{subsec:setup}}
%%%%%%%%%%%%%%%%%%%%%%%%%%%%%%%%%%%%%%%%%%%%%%%%

Let us first specify the setup for our numerical calculations.
We assume that the prefactors $\alpha_s S_\perp$ in the formulas
\eqref{eq:itmd-large-Nc} and \eqref{eq:cgc-large-Nc} are common
constants and cancel out when we take a ratio of these dijet cross-sections.
For the collinear gluon distribution $f_{g/p}$ on the projectile side,
we use the parametrization CTEQ6M\,\cite{Pumplin:2002vw}
with the factorization scale set to $\mu=(\absol{\bm p_{1t}}+\absol{\bm p_{2t}})/2$.

For the small-$x$ gluons $F(x_2,k_t)$ on the dense target side, we include the $x$-evolution effects by adopting a numerical solution
to the BK equation\,\cite{Balitsky:1995ub,Kovchegov:1999yj}:
\begin{align}
-\frac{d S_{\rm BK}({\bm r_\perp};x_2)}{d\ln(1/x_2)}
= \int d^2 {\bm r_{1\perp}} \mathcal{K}({\bm r_\perp}, {\bm r_{1\perp}};\alpha_s) 
\left[  S_{\rm BK}({\bm r_\perp};x_2) - S_{\rm BK}({\bm r_{1\perp}};x_2)S_{\rm BK}({\bm r_{2\perp}};x_2)\right]
\, ,
\end{align}
where
$Y\equiv \ln(1/x_2)$ is the evolution rapidity,
${\bm r}_\perp = {\bm r}_{1\perp} + {\bm r}_{2\perp}$ the size of
a parent dipole.
In the Gaussian truncation, $S_{q\bar q}=( S_{\rm BK} )^{1\!-\!1/N_c^2}$\,\cite{Marquet:2007vb}, therefore in the large-$N_c$ we simply use $S_{\rm BK}$
to obtain the Fourier transform $F$ using \eqref{eq:FTofSqq}.
The possible impact parameter dependence of the dipole amplitude is neglected here.
We employ the kernel $\mathcal{K}$ with running coupling corrections, which was  
derived in Ref.\,\cite{Balitsky:2006wa},
and we adopt the one-loop running coupling constant
in coordinate space
$\alpha_s(r_\perp^2)= \left [\frac{9}{4\pi} \ln \left (\frac{4C^2}{r_\perp^2\Lambda^{\prime2}}+a \right ) \right ]^{-1}$
with $C=1$.
The parameter $a$ is a smooth cutoff to make the coupling
finite in the large-dipole limit:
$\alpha_s(\absol{\bm r_\perp} \to \infty)=0.5$\,\cite{Fujii:2013gxa,Ma:2017rsu}. 
Our result on dijet production here is insensitive to this particular
choice.

For our purpose of ITMD/CGC comparison,
we take as the initial condition of the BK equation 
the McLerran-Venugopalan (MV) type
model\,\cite{McLerran:1993ka,McLerran:1993ni}
of the form:
\begin{align}
S_{\rm BK}({\bm r_\perp};x=x_0)=
\exp\left[-\frac{r_\perp^2Q_{0}^2}{4}\ln\left(\frac{1}{\absol{\bm r_\perp}\Lambda} + e\right)\right]\, ,
\label{eq:BK-IC}
\end{align}
where $x=x_0$ denotes the start of the small-$x$ evolution, which we take equal to $0.01$.
Other parameters are set as $Q_{0,p}^2=0.2\GeV^2$ and $\Lambda=0.241\GeV$
for the proton target, as indicated by global fitting analysis of deep inelastic scattering
small-$x$ data with the BK equation\,\cite{Albacete:2010sy,Albacete:2012xq,Lappi:2013zma}.

Figure\,\ref{fig:gtmds} displays the gluon TMDs,
$\mathcal{F}_{gg}(x_2,{\bm k_t})$ and
$\mathcal{F}_{\rm adj}(x_2,{\bm k_t})$,
obtained by solving the BK equation with the MV initial condition.
The BK evolution contains two competitive effects,
the gluon branching and merging, which results in
the increase (decrease) of the gluon distributions
in high (low) $\absol{\bm k_t}$ region
with decreasing $x_2$. In the following, we define the saturation scale $Q_s(x_2)$ by the peak position of the gluon TMD,
$\mathcal{F}_{\rm adj}(x_2,{\bm k_t})$, as a function of $\absol{\bm k_t}$ for fixed $x_2$, which is indicated with vertical dash-dotted arrows in Fig.\,\ref{fig:gtmds} (hence $Q_s(x_0)\simeq 0.5\GeV$ is slightly different from $Q_{0}$).
The $Q_s(x_2)$ value increases as $x_2$ decreases from $x_2=10^{-2}$, $10^{-4}$ to $10^{-6}$.
Those results are consistent with previous studies\,\cite{vanHameren:2016ftb,Albacete:2018ruq}.

%%%%%%%%%% nuclear target %%%%%%%%%%%%%%%%%%%%%%
For a heavy nuclear target, we replace the initial $Q_0$ value at $x=x_0$ by 
\begin{align}
Q_{0,A}^2=cA^{1/3}\,Q_{0,p}^2=\hat{c}\,Q_{0,p}^2 \,,
\end{align}
where we have introduced a parameter $c$\,\cite{vanHameren:2014lna}.
In Ref.\,\cite{Dusling:2009ni} it is shown that
$c\approx 0.25 - 0.5$ yields a reasonable fit to the nuclear structure function
$F_{2,A}(x,Q^2)$ at $x=0.0125$ measured by New Muon Collaboration.
Indeed, the CGC model calculation with a smaller value of
$\hat{c}\sim 3$ ($c \sim 0.5$)
has resulted in more reasonable description of forward heavy-flavor
production as well as quarkonium production in $p+A$
collisions\,\cite{Fujii:2015lld,Watanabe:2016ert,Fujii:2017rqa,Ma:2015sia,Ma:2017rsu}
compared to the early predictions
with $\hat{c}= 4 \textrm{--} 6$\,\cite{Fujii:2013gxa,Fujii:2013yja}.
In this paper, we choose $\hat{c}$ in the range of $2\leq \hat{c} \leq 3$
for the initial saturation scale in heavy nuclei,
Pb ($A=208$) and Au ($A=197$).

\subsection{Kinematics}{\label{subsec:kinematics}}

The total and relative momenta squared, \eqref{eq:ktglue} and \eqref{eq:Pt-def},
of the quark at $y_1$ with ${\bm p_{1t}}$ and the antiquark at $y_2$ with ${\bm p_{2t}}$ read,
respectively, 
\begin{align}
\absol{\bm k_t}^2 &= \absol{\bm p_{1t}}^2 + \absol{\bm p_{2t}}^2 + 2 \absol{\bm p_{1t}}\absol{\bm p_{2t}}\cos \phi
\geq (\absol{\bm p_{1t}}-\absol{\bm p_{2t}})^2\, ,\\
\absol{\bm P_t}^2 
&= \frac{\absol{\bm p_{2t}}^2 \absol{\bm p_{1t}}^2}{( \absol{\bm p_{1t}} e^{y_1} + \absol{\bm p_{2t}} e^{y_2})^2}\, 
\left( e^{2y_1} + e^{2y_2}  -2  e^{y_1+y_2}  \cos \phi \right)\, ,
\end{align}
where $\phi$ is the azimuthal angle between ${\bm p_{1t}}$ and ${\bm p_{2t}}$.

For definiteness, 
we set $\absol{\bm p_{1t}}=\absol{\bm p_{2t}}=\absol{\bm p_t}$ and then
\begin{align}
\absol{\bm k_t}^2 &= 2 \absol{\bm p_{t}}^2(1+\cos \phi)\,.
  \label{eq:kt}
\end{align}
By changing the azimuthal angle $\phi$, we can scan $\absol{\bm k_t}$ values
from $Q_s \sim \absol{\bm k_t} \ll \absol{\bm p_t}$ to $Q_s \ll \absol{\bm k_t} \sim \absol{\bm p_t}$.
The relative momentum $\absol{\bm P_t}$ depends on the rapidities, $y_{1,2}$.
For the equal rapidity, $y_1=y_2=y>0$,
it simplifies to 
\begin{align}
\absol{\bm P_t}^2 = \frac{\absol{\bm p_{t}}^2}{2}(1-\cos \phi) \label{eq:Pt-same}\, ,
\end{align}
and
for the rapidities with a large gap, $y_2 \gg y_1 >0$,
it is approximated by 
\begin{align}
\absol{\bm P_t}^2 \sim \absol{\bm p_{t}}^2
  \left (1-2e^{-(y_2-y_1)}(1+\cos \phi)\right )
  \, ,
  \label{eq:Pt-gap}
\end{align}
which is almost independent of $\phi$ as $e^{-(y_2-y_1)} \ll 1$.

%%%%%%%%%%%%%%%%%%%%%%% figure %%%%%%%%%%%%%%%%%%%%
\begin{figure}[t]
\centering
\includegraphics[width=\textwidth]{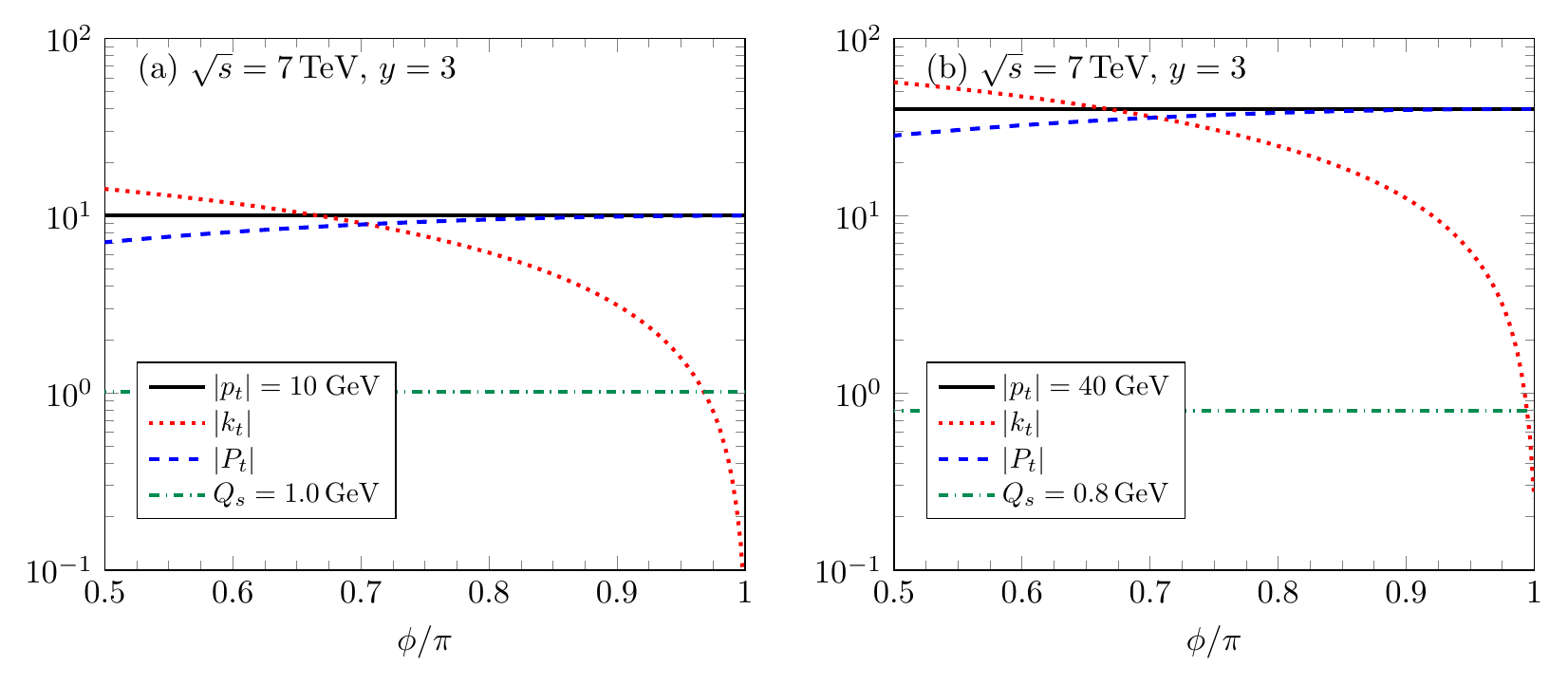}
\includegraphics[width=\textwidth]{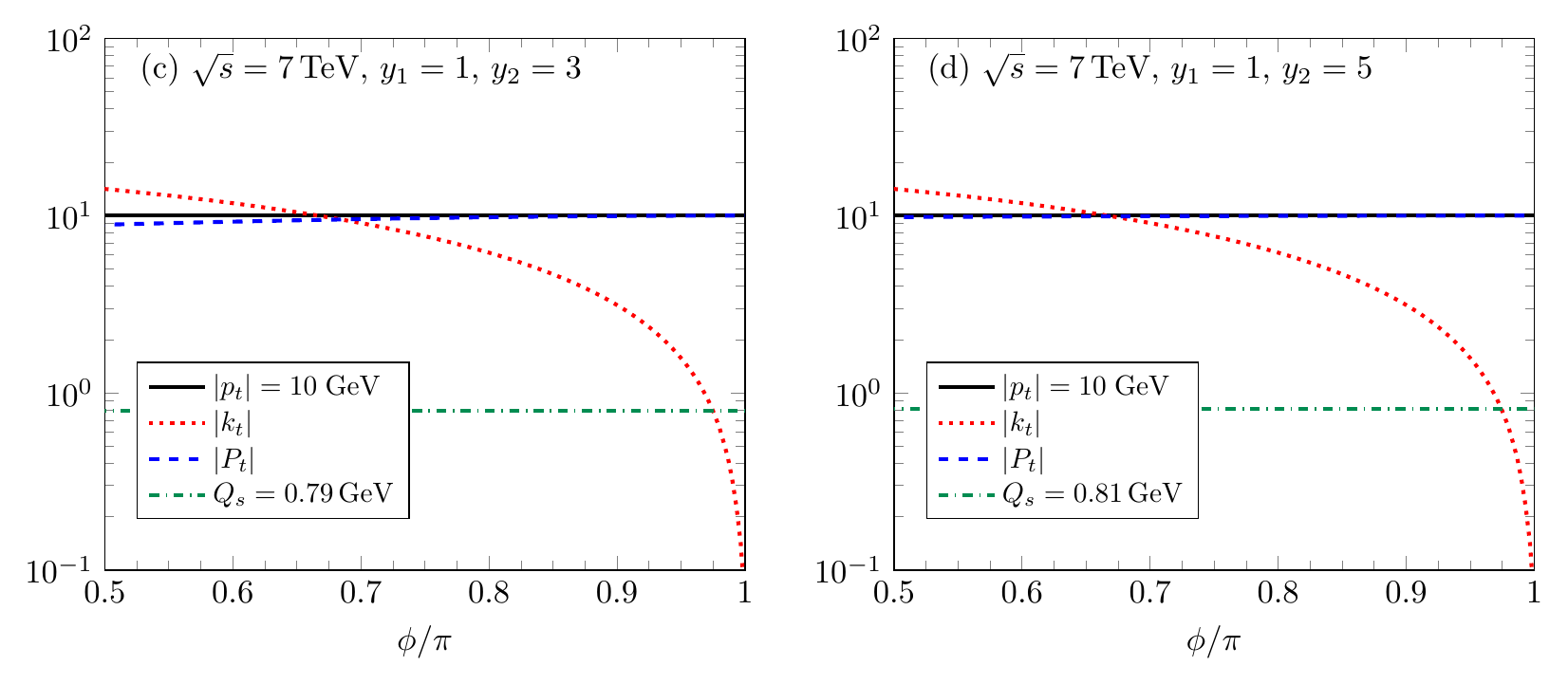}
\caption{Azimuthal angle dependence of $\absol{\bm k_t}$ (dotted) and $\absol{\bm P_t}$
  (dashed) with
  (a):\,$\absol{\bm p_t}=10\GeV$ and $y_1=y_2=3$,
  (b):\,$\absol{\bm p_t}=40\GeV$ and $y_1=y_2=3$,
  (c):\,$\absol{\bm p_t}=10\GeV$ and $y_1=1$ and $y_2=3$,
  and
  (d):\,$\absol{\bm p_t}=10\GeV$ and $y_1=1$ and $y_2=5$.
  Saturation momentum $Q_s$ (dash-dotted) in each plot is determined as the peak
  position of the gluon TMD at $\sqrt{s}=7\TeV$.
}
\label{fig:kt-Pt-Qs}
\end{figure}
%%%%%%%%%%%%%%%%%%%%%%% figure %%%%%%%%%%%%%%%%%%%%

Figure\,\ref{fig:kt-Pt-Qs} shows $\absol{\bm k_t}$ (dotted) and $\absol{\bm P_t}$ (dashed)
as a function of the azimuthal angle $\phi$ 
with fixed (a) $\absol{\bm p_t}=10\GeV$ and (b) $40\GeV$ 
for the pair at a common rapidity, $y=3$, at $\sqrt{s}=7\TeV$.
In the lower panels of Fig.\,\ref{fig:kt-Pt-Qs} shown are
the same plots but with rapidity difference,
(c) $y_1=1$ and $y_2=3$,
and
(d) $y_1=1$ and $y_2=5$,
with fixed $\absol{\bm p_t}=10\GeV$.
The $x_2$ value is fixed by Eq.\,(\ref{eq:x1x2}),
and the corresponding saturation scale $Q_s(x_2)$ 
is indicated with dash-dotted line in each plot.
The HEF formula is justified
for $Q_s\ll \absol{\bm k_t} \sim \absol{\bm P_t}$, i.e., away from the correlation limit.
On the other hand, the TMD factorization formula
applies to the kinematical region,
$ \absol{\bm k_t} \sim Q_s   \ll \absol{\bm P_t}$.

We remark here that
the ITMD formula will be less accurate due to genuine higher-twist corrections
in powers of $Q_s/ \absol{\bm p_t}$ when the separation
of the scales $\absol{\bm p_t}$ and $Q_s$ becomes marginal by lowering $\absol{\bm p_t}$,
while the CGC formula is valid as long as $Q_s(x_2) \gg \Lambda_{\mathrm{QCD}}$.

Before closing this subsection, we comment on the singularity appearing in the integrand at ${\bm q_t} - {\bm p_{2t}}={\bm 0_t}$ in the CGC formula \eqref{eq:cgc-large-Nc}.
It is not present in the ITMD formula and entirely pertains to the genuine-twists terms of Fig.\,\ref{fig:diagrams} (b).
It corresponds to the initial collinear gluon splitting into collinear quark/antiquark, which then independently pick up their transverse momentum from the two gluons (one with $p_1$ and the other with $p_2$), which indeed requires a two-body contribution at the amplitude level. In principle, this logarithmic divergence should be absorbed into a double-parton-distribution contribution not considered here\,\cite{Lappi:2012nh}. For simplicity however, in this work we regularize it by adding a small mass term
in the numerator as $1/(({\bm q_t} - {\bm p_{2t}})^2+m^2)$ in Eq.\,\eqref{eq:cgc-large-Nc},
and replace also $1/p_{1t}^2$ and $1/p_{2t}^2$ with $1/(p_{1t}^2+m^2)$ and $1/(p_{2t}^2+m^2)$ in Eq.\,\eqref{eq:itmd-large-Nc} for consistency.
We examined the $m$-dependence of our numerical results by
comparing results of $m=1\MeV \ll \Lambda_{\mathrm{QCD}}$ and $100\MeV \sim \Lambda_{\mathrm{QCD}}$. 
We found no significant change in the ratios of the dijet cross-sections of the ITMD and CGC
formulas at the LHC energy when $\absol{\bm p_t}\sim30$ or $40\GeV$. 
However, the change becomes noticeable around $\phi \sim 0$ at lower $\absol{\bm p_t}$ in both RHIC and LHC energies.
In the following calculations, we will take $m=100\MeV$ and study the region $\phi>\pi/2$.

%%%%%%%%%%%%%%%%%%%%%%%%%%%%%%%%%%%%%%%%%%%%%%%%%%%%%%%%%%
\subsection{ITMD/CGC ratio in $p+p$}{\label{subsec:ratios}}
%%%%%%%%%%%%%%%%%%%%%%%%%%%%%%%%%%%%%%%%%%%%%%%%%%%%%%%%%%

Our focus is on azimuthal angle correlation in forward
quark dijet production. We will compute the dijet yield
\begin{align}
\frac{dN(pp/pA\to q\bar{q} X)}{dy_1 dy_2 d\absol{\bm p_{1t}} d\absol{\bm p_{2t}} d\phi}
\equiv
\frac{2\pi \absol{\bm p_{1t}}\absol{\bm p_{2t}}}{S_\perp}\frac{d\sigma(pp/pA \to q\bar{q} X)}{dy_1dy_2d^2{\bm p_{1t}}d^2{\bm p_{2t}}}
\,.
\label{eq:dijet-correlation}
\end{align}
The cross-section depends on the relative angle $\phi$, not
on individual angles of $p_{1t,2t}$ due to the rotational symmetry
of the dijet production.
Note that the $q\bar{q}$ dijet yield is given by $N =\sigma_{q\bar{q}}/S_\perp$
with the assumption $S_\perp\approx \sigma_\mathrm{inel}$ between
the effective transverse area and the inelastic cross section.

Both the CGC and ITMD formulae contain the TMD and HEF limits within the appropriate kinematics, the difference between
them represents genuine higher-twist contributions, present in the CGC results but absent in the ITMD case, where only kinematical-twist contributions in $\absol{\bm k_t}/\absol{\bm P_t}$ are resumed.
In terms of the CGC formula, that difference comes from power corrections in the dipole size expansion
\footnote{Effects of higher-twists have been studied in quark-pair production in the CGC framework in Refs.\,\cite{Fujii:2005vj,Fujii:2006ab}, but with respect to the HEF or $k_t$-factorization formula only.}.

%%%%%%%%%%%%%%%%% figure %%%%%%%%%%%%%%%%%%%%%%%%%%
\begin{figure}[t]
\centering
\includegraphics[width=\textwidth]{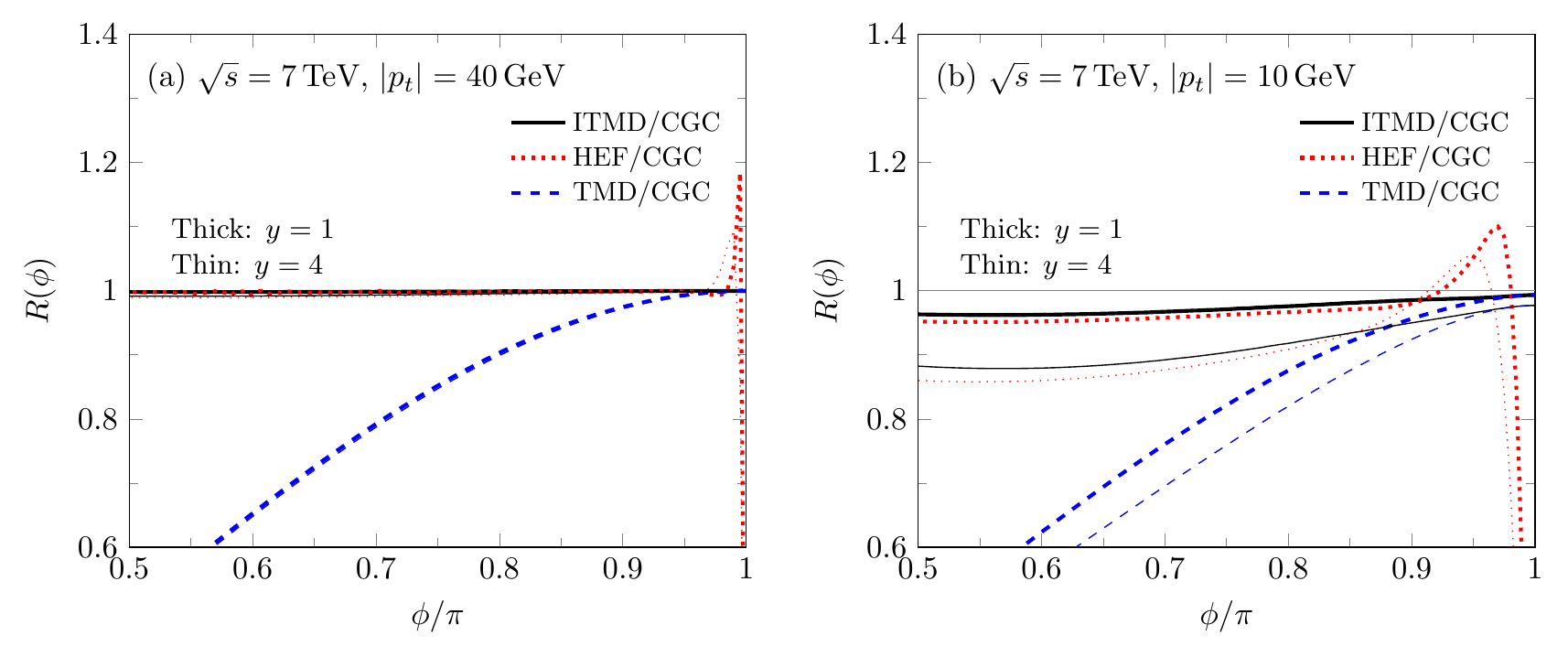}
\caption{Ratios of the ITMD to the CGC result for quark dijet production cross-section (ITMD/CGC) are 
	shown in solid black line as a function of the azimuthal angle $\phi$ between the jets at (a): $\absol{\bm p_t}=40\GeV$ and 
	(b): $\absol{\bm p_t}=10\GeV$ in $p+p$ collisions at $\sqrt{s}=7\TeV$. Thick (thin) line denotes the results at $y=1$ ($y=4$). 
	Ratios of HEF/CGC and TMD/CGC are also plotted with red dotted, and blue dashed lines, respectively.}
\label{fig:dijet-ratio-ydep}
\end{figure}

\begin{figure}[t]
\centering
\includegraphics[width=\textwidth]{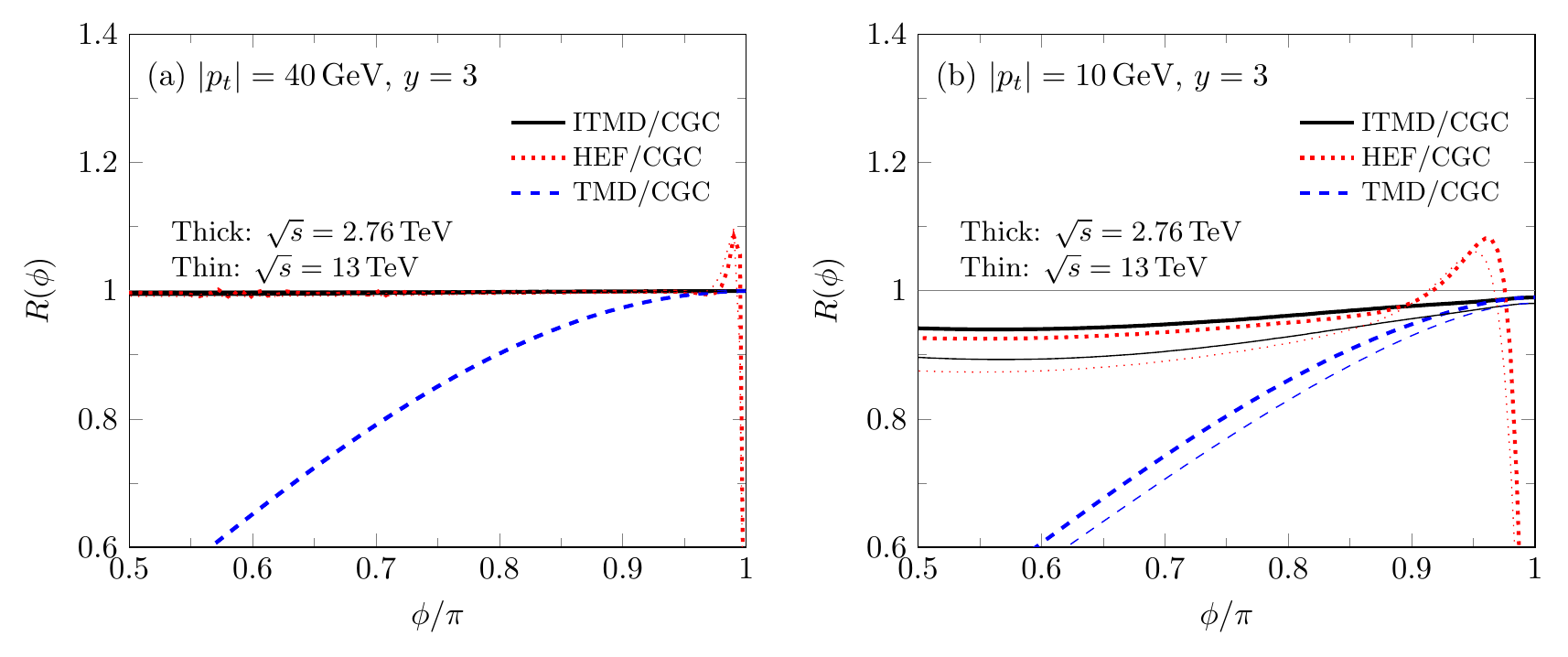}
\caption{Ratio of the ITMD/CGC ratios for quark dijet production in p+p collisions
  at $\sqrt{s}=2.76\TeV$ (thick line) and $13\TeV$ (thin line)
  for (a): $\absol{\bm p_t}=40\GeV$  and (b): $\absol{\bm p_t}=10\GeV$ with $y=3$ fixed.  
Other notations are the same in Fig.\,\ref{fig:dijet-ratio-ydep}.}
\label{fig:dijet-ratio-sdep}
\end{figure}
%%%%%%%%%%%%%%%%% figure %%%%%%%%%%%%%%%%%%%%%%%%%

In order to quantify the genuine higher-twist effects, we compare results of the ITMD formula and of the CGC
by taking the ratio of the former to the latter.
In Fig.\,\ref{fig:dijet-ratio-ydep}, we show the ratio $R$  as a
function of $\phi$ for the pair of the common rapidity $y_1=y_2=y$
at (a) $\absol{\bm p_t}=40\GeV$ and (b) $10\GeV$.
The ITMD/CGC ratio $R$ for $y=1$ (thick black line)
in Fig.\,\ref{fig:dijet-ratio-ydep} (a) is consistent with unity over
the whole range of $\phi$ studied here, and for $y=4$ it lies barely below
unity as $Q_s(x_2)$ becomes larger.
Other ratios of TMD/CGC (blue dashed) and HEF/CGC (red dotted)
deviate from unity outside of their respective domain of applicability, as is expected.
Indeed, the dijet production cross-section in the HEF formula unphysically
vanishes $d\sigma_\mathrm{HEF}({\bm k_t}\to {\bm 0_t}) \to 0$ in the back-to-back
limit (see Eq.\,\eqref{eq:HEF}).
On the other hand, the TMD formula, which ignores the off-shellness 
of the partons in the hard matrix factors,
underestimates the cross-section for $\phi$
away from the back-to-back region.

At the lower $\absol{\bm p_t}=10\GeV$ (Fig.\,\ref{fig:dijet-ratio-ydep} (b)),
the ITMD/CGC ratio $R$ shifts below unity by $\lesssim
5$ to $\lesssim 15$\% 
around $\phi \sim \pi/2$ as the
rapidity $y$ increases from $y=1$ to $4$.
This deviation can be understood as a result of the
increase of the power corrections mentioned earlier. 
We stress here that the ITMD formula approximates the CGC result uniformly
over the region of $\phi$ with 5--15\% accuracy. 
The genuine higher-twist corrections 
become more important outside the back-to-back region, while it is negligible 
around the back-to-back limit $\phi=\pi$.
We will investigate the power corrections further in the last subsection below.

%%%%%%%%%%%%%%%% figure, with rapidity gap %%%%%%%%%%%%
\begin{figure}[t]
	\centering
	\includegraphics[width=\textwidth]{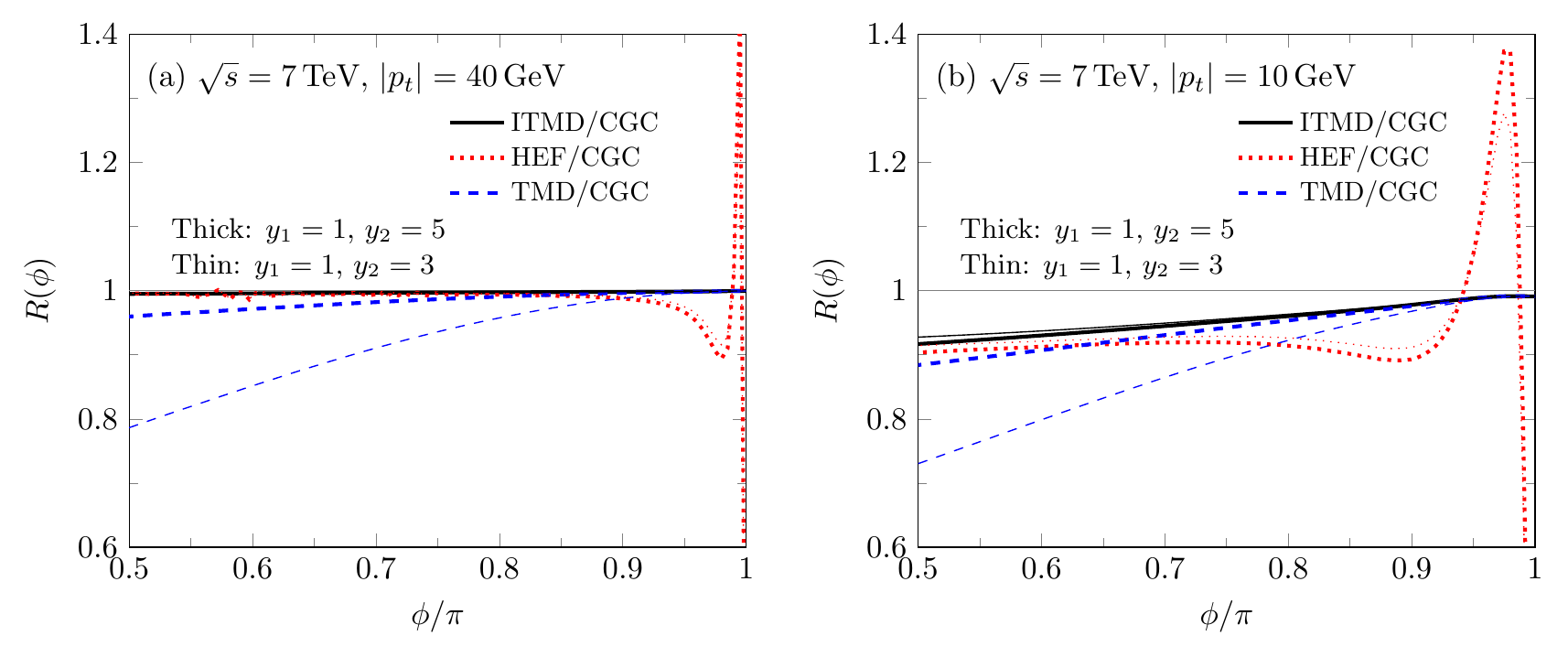}
	\caption{The ratio $R$ of ITMD/CGC (solid black line) for quark dijet
		production with $(y_1,y_2)=(1, 5)$ (thick) and $(1,3)$ (thin)
		at (a): $\absol{\bm p_t}=40\GeV$ and (b): $10\GeV$ in
		$p+p$ collisions at $\sqrt{s}=7\TeV$.
		Other notations are the same in Fig.\,\ref{fig:dijet-ratio-ydep}.}
	\label{fig:dijet-ratio-ydep2}
\end{figure}
%%%%%%%%%%%%%%%%%%%%%%%%%%%%%%%%%%%%%%%%%%%

In Fig.\,\ref{fig:dijet-ratio-sdep}, we study the energy dependence of
the ITMD/CGC ratio $R$ by showing the cases
of $\sqrt{s}=2.76\TeV$ (thick) and $13\TeV$ (thin) for (a) $\absol{\bm p_t}= 40\GeV$ and (b) $10 \GeV$.
For the larger $\absol{\bm p_t}$ (a), we find that the ITMD approximation is very accurate there,
almost the same result as in Fig.\,\ref{fig:dijet-ratio-ydep},
which indicates that the corrections in $Q_s/p_t$ are well suppressed there.
For the lower $\absol{\bm p_t}$ (b), the ITMD/CGC ratio deviates from unity and  
the depletion becomes more significant with
increasing collision energy $\sqrt{s}=2.76$ to
$13\TeV$ (i.e., increasing $Q_s^2(x_2)$).

We also examine dijet production with a rapidity separation
in the cases, $(y_1, y_2)=(1,3)$ and (1,5),
as shown in thick and thin lines, respectively,
in Fig.\,\ref{fig:dijet-ratio-ydep2}
for (a) $\absol{\bm p_t}=40\GeV$  and (b) $10\GeV$. 
From Fig.\,\ref{fig:dijet-ratio-ydep2} (a), we see that the ITMD formula
well approximates the CGC result, and 
interpolates the results of TMD and HEF formulas uniformly over the
range of $\pi/2 \lesssim \phi \lesssim \pi$.
The TMD estimate becomes accurate near the back-to-back region
$\phi \sim \pi$, while it is less accurate for $\phi$ away from it.
For larger $y_2$, this approximation becomes better.
On the other hand, unsurprisingly, the HEF formula fails to reproduce the CGC
result in a wider region of $\phi$ around $\sim \pi$
irrespective of the $y_2$ value.

At the lower $\absol{\bm p_t}=10\GeV$ (Fig.\,\ref{fig:dijet-ratio-ydep2} (b)),
the ITMD formula interpolates still smoothly from the TMD to the HEF
results with decreasing $\phi$ from the back-to-back limit $\phi\sim \pi$.
However, the value of the ITMD/CGC ratio lies significantly below unity
in the non-back-to-back region, which reflects
the size of the genuine twist corrections.

%%%%%%%%%%%%%%%%% figure %%%%%%%%%%%%%%%%%%%%%%%%%%

\begin{figure}[t]
\centering
\includegraphics[width=\textwidth]{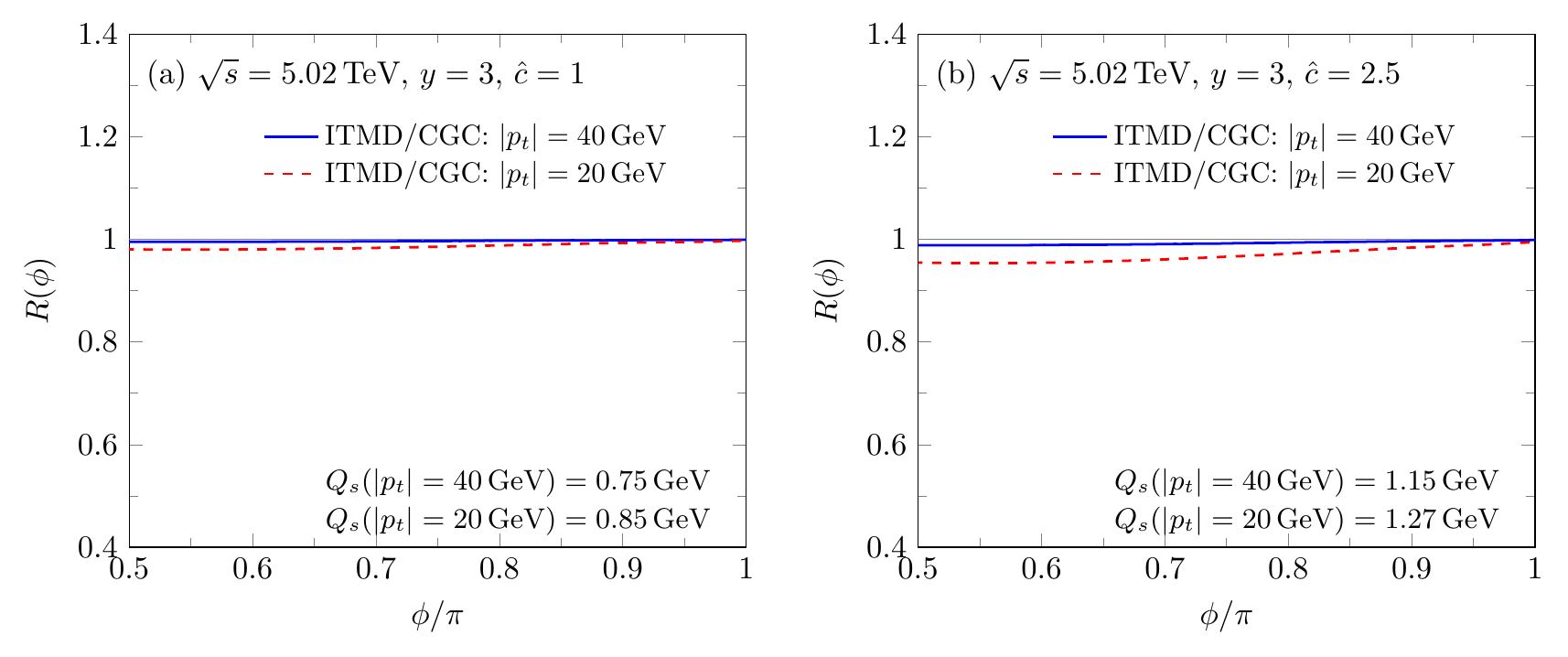}
\includegraphics[width=\textwidth]{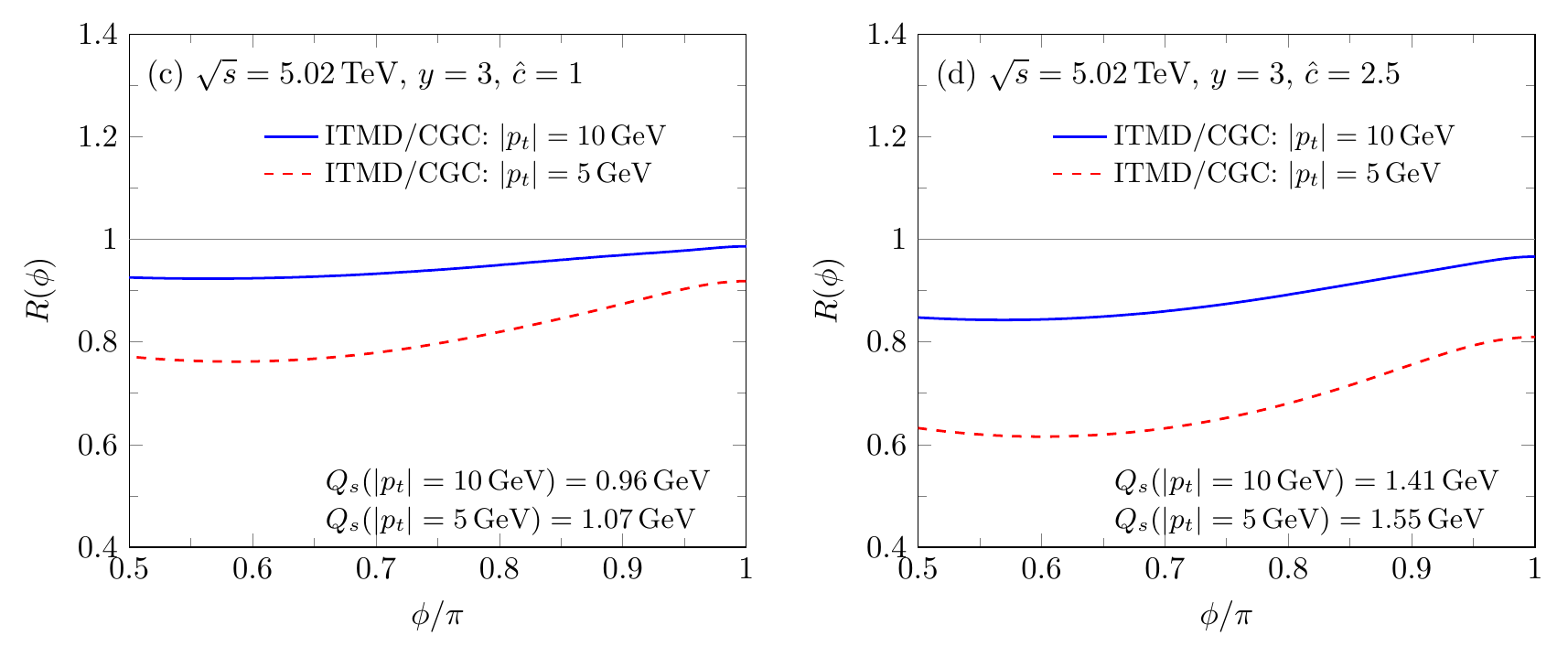}
\caption{The ITMD/CGC ratio $R$ in $p+p$ 
  collisions (left; $\hat c=1$) and in $p+A$ collisions (right; $\hat c=2.5$) at
  $y=3$ at $\sqrt{s}=5.02\TeV$.
  (a):\, Results for $\absol{\bm p_t}=40$ (blue solid) and $20\GeV$ (red dashed) in $p+p$.
  (b):\, The same as in (a) but in $p+A$.
  (c):\, Results for $\absol{\bm p_t}=10$ (blue solid) and $5\GeV$ (red dashed) in $p+p$.
  (d):\, The same as in (c) but in $p+A$.
  The $Q_s$ value determined by the gluon TMDs is shown in each panel.}
\label{fig:dijet-ratio-pA-lhc}
\end{figure}
%%%%%%%%%%%%%%%%% figure %%%%%%%%%%%%%%%%%%%%%%%%%%

%%%%%%%%%%%%%% figure: at RHIC energy %%%%%%%%%%%%%%%%%%
\begin{figure}[t]
\centering
\includegraphics[width=\textwidth]{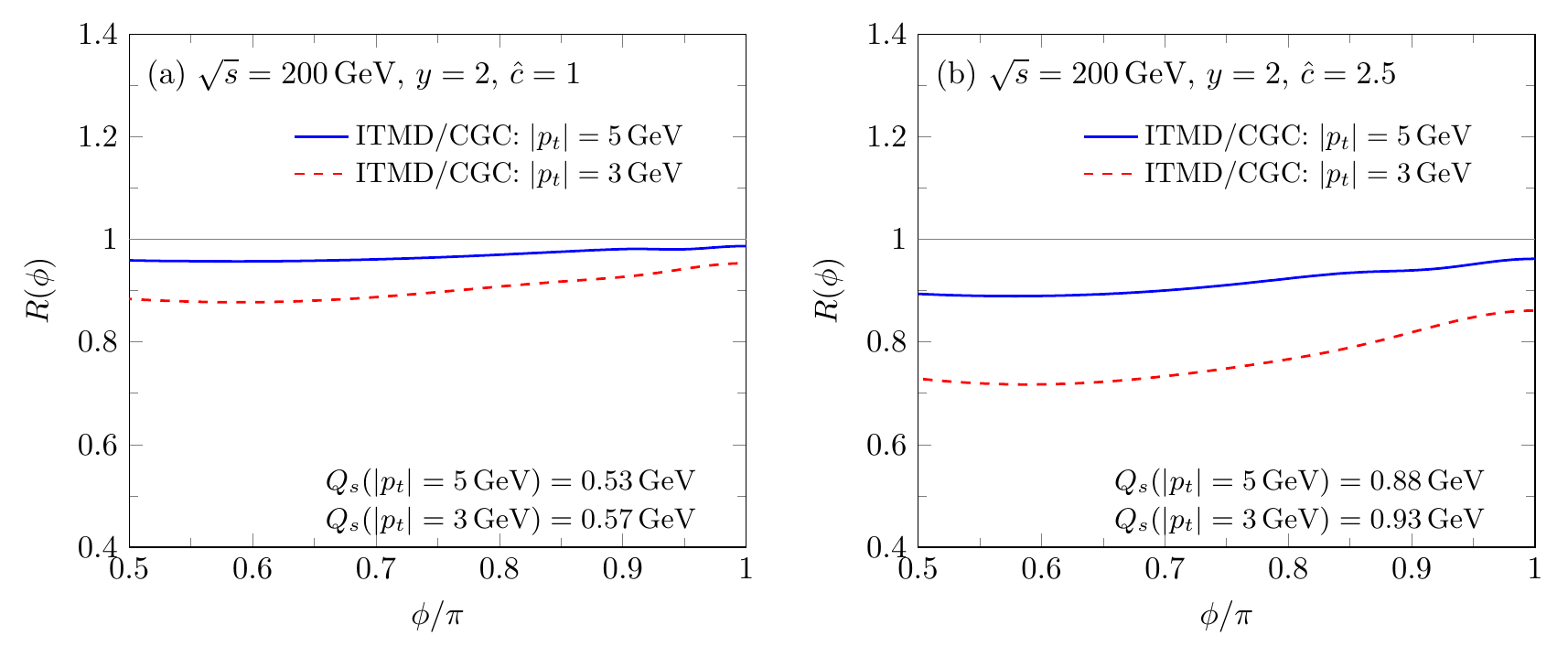}
\caption{Comparison of the ITMD/CGC ratio $R$ in (a) $p+p$ collisions
and (b) $p+A$ collisions at $y=2$ at $\sqrt{s}=200\GeV$.
The results for $\absol{\bm p_t}=5\, (3)\GeV$ are shown in blue
solid (red dashed) lines.
The $Q_s$ value determined by TMDs is shown in each panel.}
\label{fig:dijet-ratio-pA-rhic}
\end{figure}
%%%%%%%%%%%%%% figure: at RHIC energy %%%%%%%%%%%%%%%%%%

\subsection{ITMD/CGC ratio in $p+A$ and nuclear modification factor}
           {\label{subsec:nuclear-effect}}

The saturation scale $Q_{sA}^2$ in a heavy nucleus will be enhanced
by a factor of ${\hat c}\propto A^{1/3}$ compared to $Q^2_{sp}$,
as discussed in Sec.\,\ref{subsec:setup}, and therefore it is valuable to analyze the nuclear dependence of
the ITMD/CGC ratio in forward dijet production in $p+A$ collisions.
We plot in Fig.\,\ref{fig:dijet-ratio-pA-lhc} the ratios $R$
in $p+p$ ($\hat c=1$) and $p+A$ ($\hat c=2.5$) collisions
at $\sqrt{s}=5.02\TeV$ for $\absol{\bm p_t}=40$ and $20\GeV$ ((a) and (b)),
and for $\absol{\bm p_t}=$ 10 and $5\GeV$ ((c) and (d)).

From the comparison of the $R$ ratios in $p+p$ and $p+A$ collisons at
$\absol{\bm p_t}=40$ and $20\GeV$ in Fig.\,\ref{fig:dijet-ratio-pA-lhc} (a) (b),
we find that the deviation of the ratio $R$ from unity becomes
more noticeable in $p+A$ collisions and for the lower $\absol{\bm p_t}=20\GeV$,
which indicates the enhanced power corrections of $Q_s/\absol{\bm p_t}$
at lower $\absol{\bm p_t}$.
At yet lower values $\absol{\bm p_t}=10,\, 5\GeV$, the deviation becomes significant even in $p+p$ case
(Fig.\,\ref{fig:dijet-ratio-pA-lhc} (c)),
and is more profound in $p+A$ case (Fig.\,\ref{fig:dijet-ratio-pA-lhc} (d)).
In these cases,
the ITMD is no longer a good approximation to the CGC.
The genuine twist corrections do not vanish even in the correlation limit, so long as $\absol{\bm p_t}$ is a finite value much bigger than $Q_s$\,\footnote{One can verify analytically that $1-R\sim Q_s^2/p_t^2$ by using the GBW gaussian model for $F$.}.

Figure\,\ref{fig:dijet-ratio-pA-rhic} shows the results at the RHIC
energy, $\sqrt{s}=200\GeV$.  
Since the dijet production formulas,
Eqs.\,\eqref{eq:itmd-large-Nc} and \eqref{eq:cgc-large-Nc}
premise that $x_2$ is small, $x_2\leq x_0=0.01$,
the jet momentum $\absol{\bm p_t}$ is accordingly limited to the lower values,
and here we take $y=2$ and $\absol{\bm p_t}=5$ and $3\GeV$.
Although $Q_s$ becomes smaller at RHIC,
the ratios $R$ for $\absol{\bm p_t}= 5$ and $3\GeV$ at $\sqrt{s}=200\GeV$
in Fig.\,\ref{fig:dijet-ratio-pA-rhic}
deviate from unity similarly to those for $\absol{\bm p_t}=10$ and $5\GeV$ at
$\sqrt{s}=5.02\TeV$ in Figs.\,\ref{fig:dijet-ratio-pA-lhc} (c) (d).
This result shows the dijet production at the RHIC energy is sensitive to
genuine higher-twist corrections in $Q_s/\absol{\bm p_t}$, which are
not included in the ITMD formula.

%%%%%%%%%%%%%%%%%% figure: Nuclear modification factors %%%%%%%%%%%%%%%%%

\begin{figure}[t]
\centering
\includegraphics[width=0.49\textwidth]{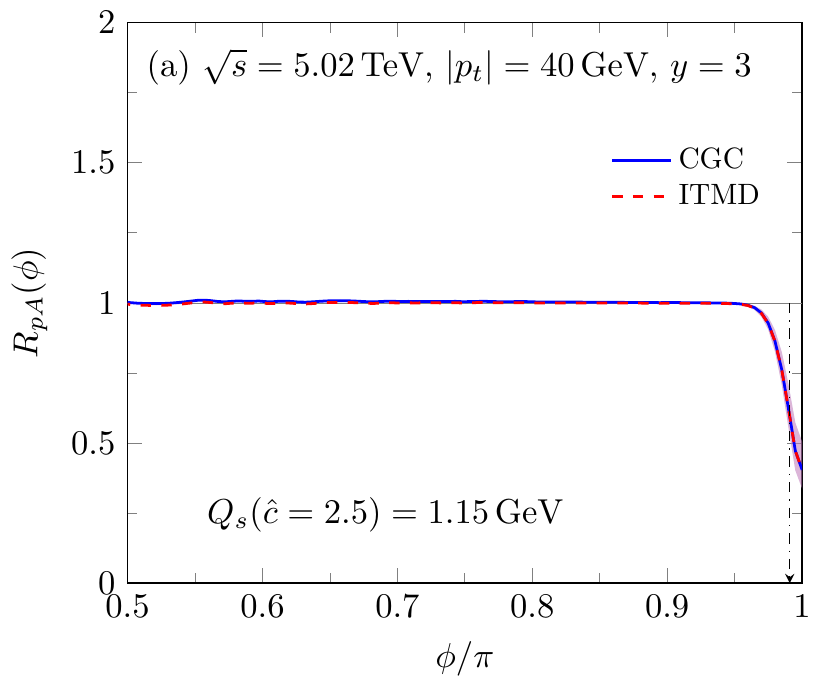}
\includegraphics[width=0.49\textwidth]{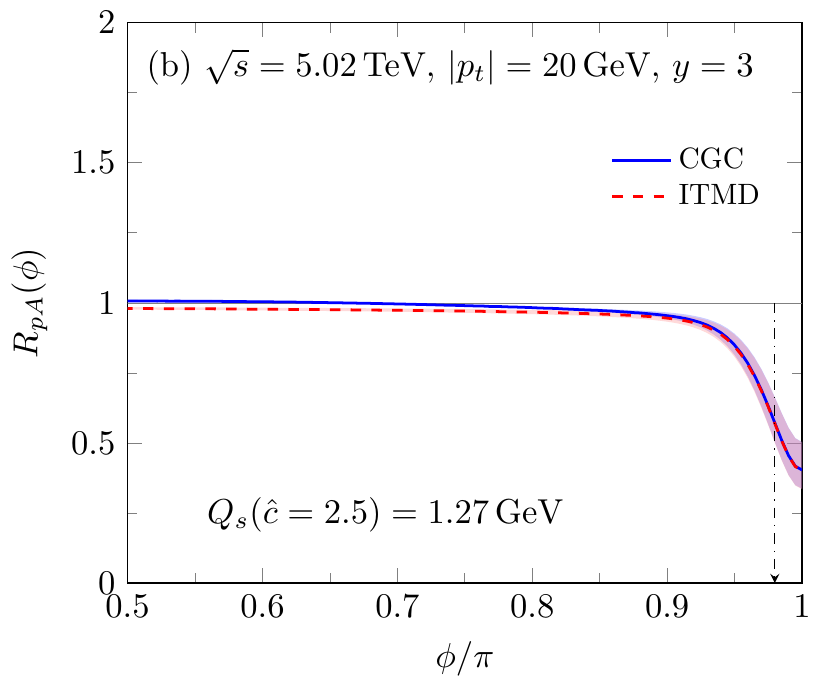}
\includegraphics[width=0.49\textwidth]{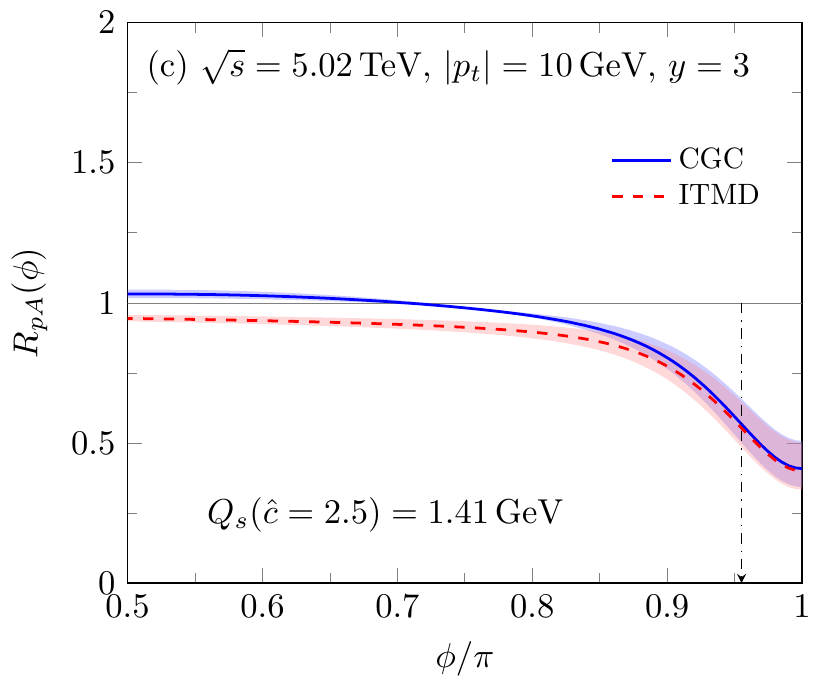}
\includegraphics[width=0.49\textwidth]{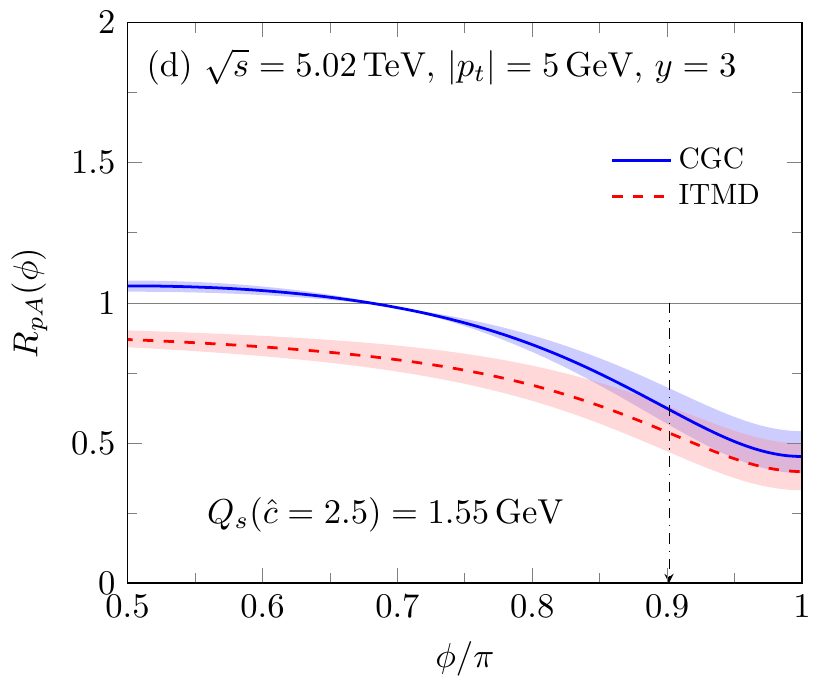}
\caption{Nuclear modification factor as a function of $\phi$ for forward dijet production of 
$\absol{\bm p_t}=40\GeV$ (a), $20\GeV$ (b), $10\GeV$ (c), and $5\GeV$ (d) at $\sqrt{s}=5.02\TeV$. 
Colored bands show the uncertainty of the initial saturation scale for the target nucleus: $\hat{c}=2\textrm{--}3$. 
The vertical arrow line in each plot indicates the deviation $\delta\phi=Q_s/\absol{\bm p_t}$ from the correlation limit $\phi=\pi$.}
\label{fig:dijet-RpA-lhc}
\end{figure}

\begin{figure}[t]
\centering
\includegraphics[width=0.49\textwidth]{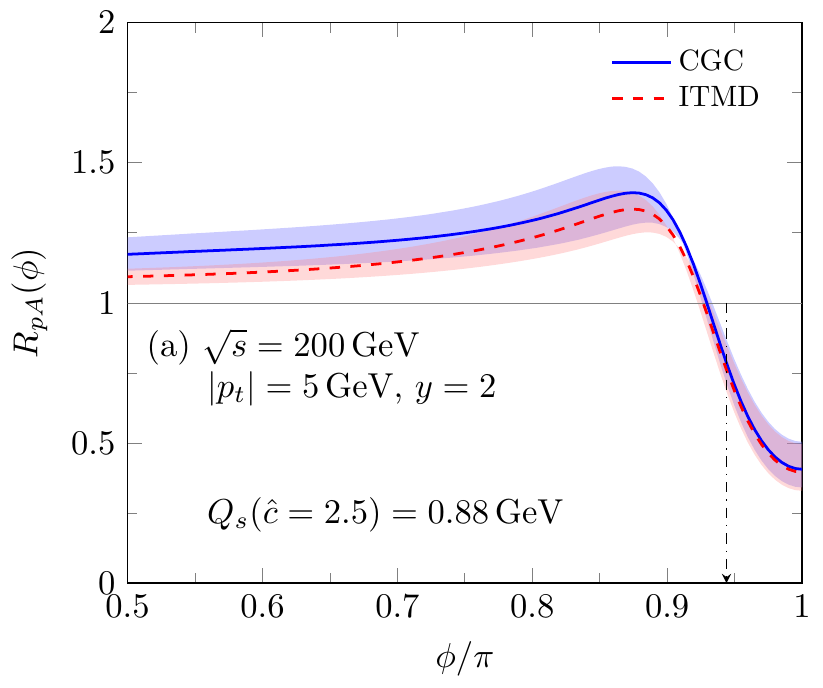}
\includegraphics[width=0.49\textwidth]{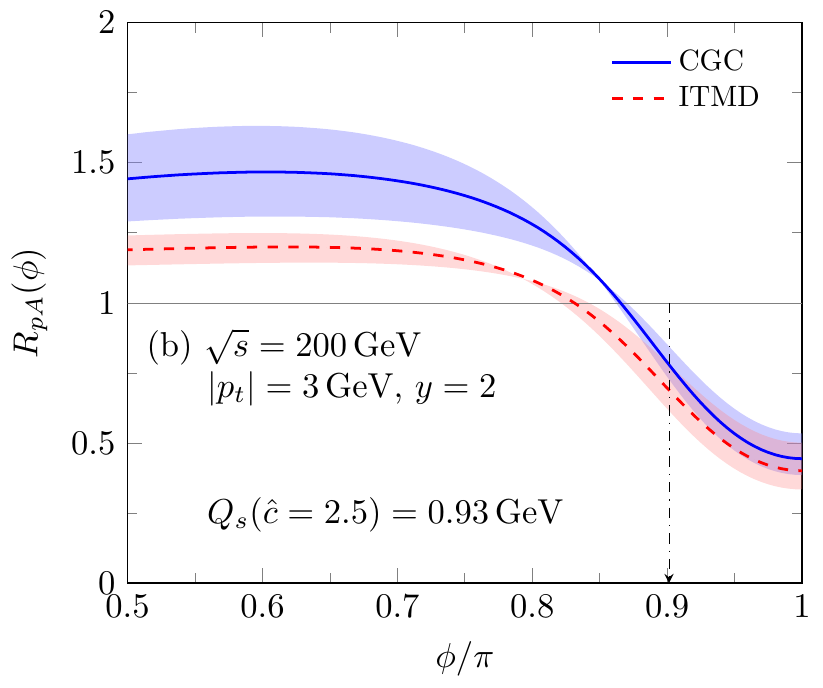}
\caption{Nuclear modification factor as a function of $\phi$ for forward dijet production of 
$\absol{\bm p_t}=5\GeV$ (a) and $3\GeV$ at $\sqrt{s}=200\GeV$. Notation is the same in Fig.\,\ref{fig:dijet-RpA-lhc}.}
\label{fig:dijet-RpA-rhic}
\end{figure}

%%%%%%%%%%%%%%%%%% end of fig: Nuclear modification factors  %%%%%%%%%%%%%%%%%

Next let us discuss the so-called nuclear modification factor $R_{pA}$,
for forward dijet production ($d{\mathcal{P.S.}}=dy_1 dy_2 d\absol{\bm p_{1t}} d\absol{\bm p_{2t}} d\phi$):
\begin{align}
  R_{pA}\equiv\frac{1}{A}\frac{S_\perp^A}{S_\perp^p}
  \frac{\frac{dN(pA\to q\bar q X)}{d\mathcal{P.S.}}}
       {\frac{dN(pp\to q\bar q X)}{d\mathcal{P.S.}}}\,,
\label{eq:RpA}	
\end{align}
where $S_\perp^{p,A}$ denote the effective transverse areas of the proton
and nucleus targets, respectively.
A $p+A$ collision should be presumably regarded as a superposition of $p+p$ collisions
for the high momentum limit $\absol{\bm p_t} \to \infty$ (at $\phi \ne \pi$),
and then $R_{pA} \to 1$ is expected.
To assure this constraint, we normalize the effective transverse area
in our model calculations as
\begin{align}
\frac{1}{A}\frac{S_\perp^A}{S_\perp^p}=\frac{1}{\hat{c}}.
\end{align}
A modification of $R_{pA}$ from unity signals the presence of
nuclear effects. Figure\,\ref{fig:dijet-RpA-lhc} demonstrates $R_{pA}$ of the quark dijet
production at $y=3$ at $\sqrt{s}=5.02\TeV$
for jet momentum $\absol{\bm p_t}=40, \,20, \,10$, and $5\GeV$.
Colored bands depict the uncertainty 
estimated by the change of the results
when the initial saturation scale of the nucleus
$Q_{s0,A}^2 = \hat{c}\, Q_{s0,p}^2$ is varied
in the range of $\hat c=2\textrm{--}3$.

At $\absol{\bm p_t}=40\GeV$ (Fig.\,\ref{fig:dijet-RpA-lhc} (a)),
the CGC (blue solid), and ITMD (red dashed) formulas give
the same prediction for $R_{pA}$.
The prediction is consistent with unity over a wide range of
$\phi$ and is suppressed only in the TMD regime in the vicinity of $\phi = \pi$, where
the total transverse momentum of the dijet becomes small and comparable to the saturation scale: $\absol{\bm k_t} \lesssim Q_{s}(x_2)$.
The intrinsic transverse momentum of the gluons, which is of the order of $Q_{s}$, is larger in the heavy nucleus than 
in the proton and smears the azimuthal angle correlation.
The region of the suppression should be characterized by
$\delta\phi= |\pi - \phi| \lesssim Q_{sA}/\absol{\bm p_t}$
which we indicate with a vertical dash-dotted
arrow in Fig.\,\ref{fig:dijet-RpA-lhc} (a). Decreasing the jet momentum $\absol{\bm p_t}$ from (a) $40\GeV$ down to (d) $5\GeV$
in Fig.\,\ref{fig:dijet-RpA-lhc},
we find that the suppression of $R_{pA}$ appears in a wider range of $\phi$
on the away side.
This is because, for lower $\absol{\bm p_t}$ at fixed $y$,
the relevant $x_2$ is smaller and $Q_s(x_2)$ is larger accordingly,
and therefore the region of $\delta \phi \lesssim Q_s/\absol{\bm p_t}$ gets wider.

We also notice that the difference between the ITMD and CGC results
increases as $\absol{\bm p_t}$ decreases. In the suppression regime, near $\phi=\pi$,
the differences stay rather small, but at moderate $\phi$ away from $\phi \sim \pi$,
those differences can get large and in fact, in that regime the CGC formula exceeds
unity while the ITMD one stays suppressed. This qualitative change is caused by the
genuine higher-twist corrections (Fig.\,\ref{fig:diagrams} (b)) present in the CGC formula. They contribute significantly to quark dijet production
at moderate values of $\phi$ (where the ITMD cross-section is not very large), and contribute even more so in $p+A$ collisions compared to $p+p$ collisions,
due to the bigger saturation scale in the former case. That creates an enhancement of $R_{pA}$.

%%%%%%%%%%%%%%% figure RpA vs F_gg %%%%%%%%%%%%%%%
\begin{figure}[t]
\centering
	\includegraphics[width=\textwidth]{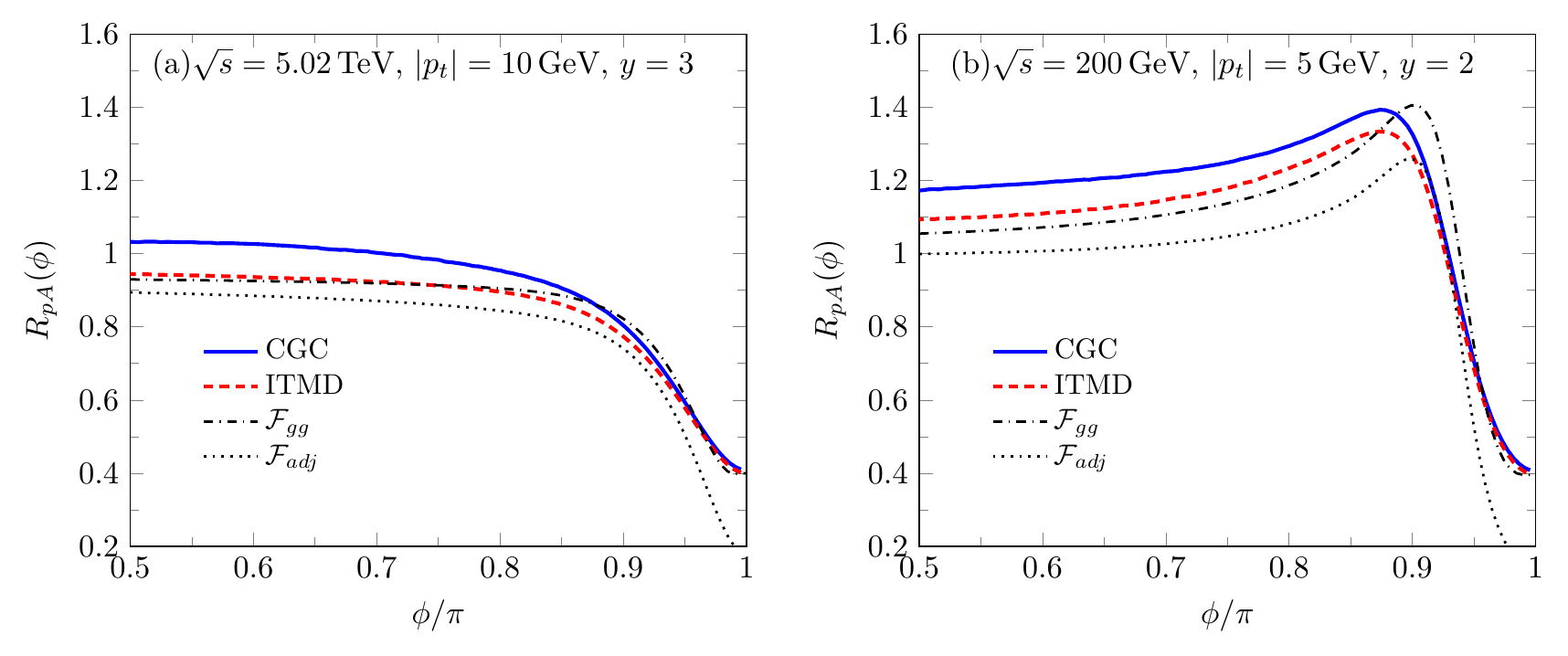}
	\caption{(a): $R_{pA}$ obtained in the CGC
  (blue solid), ITMD (red dense dashed) formulas
 for $q\bar q$ dijet production with $\absol{\bm p_t}=10\GeV$ 
 and $y=3$ in $p+A$ collisions ($\hat c=2.5$) at $\sqrt{s}=5.02\TeV$.
 For comparion, $R_{pA}$ of the gluon TMDs with $\hat{c}=2.5$ to those with $\hat{c}=1$,
 devided by $\hat{c}=2.5$, is shown for
 $\mathcal{F}_{gg}$ (black dash-dotted) and
 $\mathcal{F}_{\rm adj}$ (black dotted).
 (b): The same with $\absol{\bm p_t}=5\GeV$ and $y=2$ at $\sqrt{s}=200\GeV$.
        }
	\label{fig:power-ratio}
\end{figure}

%%%%%%%%%%%%%%%%%%%%%%%%%%%%%%%%%%%%%%%%%%%%%%%%%%%%%%%%%%%

At the RHIC energy $\sqrt{s}=200\GeV$, 
a similar suppression of $R_{pA}$ is seen in the back-to-back region
around $\delta \phi < Q_{sA}/\absol{\bm p_t}$ in Fig.\,\ref{fig:dijet-RpA-rhic},
reflecting the larger intrinsic ${\bm k_t}$ in the nuclear target. 
In contrast, at $\phi$ away from $\pi$, both the CGC and ITMD results show enhancements of $R_{pA}$.
The larger discrepancy between the ITMD and CGC results in Fig.\,\ref{fig:dijet-RpA-rhic} (b) than in (a)
is again a manifestation of the larger genuine twist corrections to dijet production at $\absol{\bm p_t}=3\GeV$ compared to $\absol{\bm p_t}=5\GeV$.
The enhancement of the ITMD cross section at $\phi$ away from $\pi$ is not surprising actually; at RHIC energies we are simply sensitive to our initial conditions:
if we plot the ratio of the gluon TMD ${\cal F}_{gg}(x_2,{\bm k_t})$ of the heavy nucleus to that of the proton at $x_2\sim x_0$,
it does show a Cronin-like peak structure as a function of $\absol{\bm k_t}$. See Fig.\,\ref{fig:power-ratio}.
Indeed, Fig.\,\ref{fig:power-ratio} compares $R_{pA}$ of the quark dijet
production cross-section obtained with the CGC and ITMD formulas,
together with $R_{pA}$ of the gluon TMDs $\mathcal{F}_{gg}$
and $\mathcal{F}_{\rm adj}$ (in order to highlight the higher-twist effects, we choose the lower values of the jet momentum
$\absol{\bm p_t}=10$ and $5\GeV$, resp.\ at the LHC (a) and RHIC (b) energies).
We see that the cross-section ratio obtained with the ITMD formula is roughly proportional
to the ratio of $\mathcal{F}_{gg}$ (and, away from $\phi=\pi$, to $\mathcal{F}_{\rm adj}$ also, when that TMD is no more proportional to $k_t^2$, see Fig.\,\ref{fig:gtmds}).

%%%%%%%%%%%%%%% figure dN vs |kt|/|Pt| %%%%%%%%%%%%%%%
\begin{figure}[t]
\centering
\includegraphics[width=\textwidth]{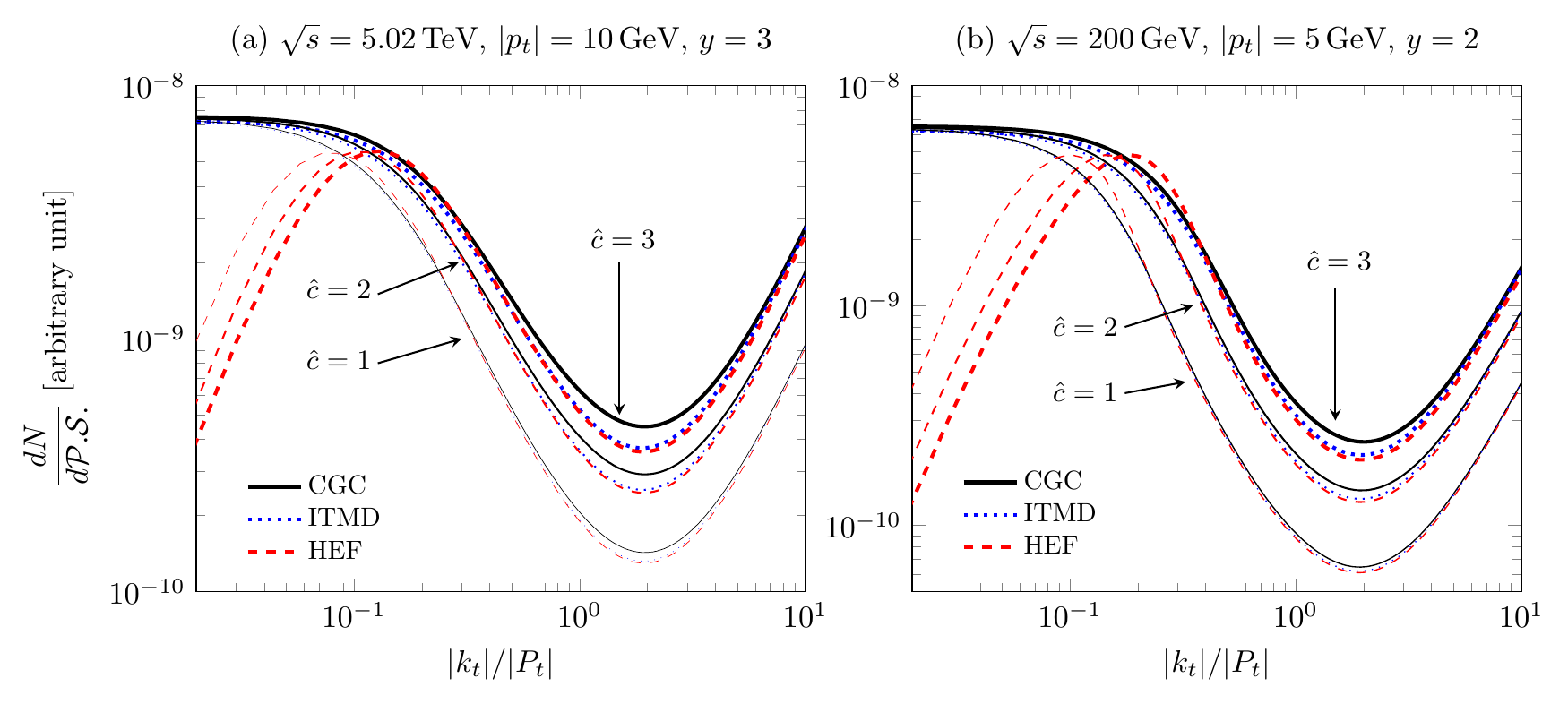}
\caption{Dijet production yield vs. $\absol{\bm k_t}/\absol{\bm P_t}$ obtained in the CGC (black solid), 
ITMD (blue dotted), and HEF (red dashed) framework at $\absol{\bm p_t}=20\GeV$ (a) and $10\GeV$ (b) with $y=3$ and $\sqrt{s}=5.02\TeV$ fixed.
Line thickness represents the nuclear dependence: $\hat{c}=3$ (very thick), $\hat{c}=2$ (semi thick), $\hat{c}=1$ (thin).
}
\label{fig:kt-dep}
\end{figure}

%%%%%%%%%%%%%%%%%%%%%%%%%%%%%%%%%%%%%%%%%%%%%%%%%%%%%%%%%%%

The contribution of multi-body scattering diagrams in the CGC to higher-twist corrections was also addressed in Refs.\,\cite{Fujii:2005vj,Fujii:2006ab}, as $k_t$-factorization breaking effect for quark-antiquark pair production. In that analysis, what was studied was the difference between the CGC and HEF formulae, which contains two types of HEF (or $k_t$) factorization breaking contributions: leading-twist saturation corrections in $Q_s/\absol{\bm k_t}$ and genuine-twist saturation corrections in $Q_s/\absol{\bm p_t}$. In the present work, by employing the ITMD framework, we are now able to include the former in the baseline, and isolate the latter as the difference between the CGC and ITMD formulae. To illustrate our findings, Fig.\,\ref{fig:kt-dep} displays the dijet production yield at the LHC and RHIC as a function of $\absol{\bm k_t}/\absol{\bm P_t}$. We find that at small values of $\absol{\bm k_t}$ (around $\absol{\bm k_t}/\absol{\bm P_t}=\mathcal{O}(0.1)$ or smaller), the leading-twist saturation corrections are responsible for the (rather large) difference between the CGC and HEF curves, as the genuine-twist corrections are negligible (since the ITMD and CGC curves coincide). By contrast, the genuine-twist saturation corrections become visible when $\absol{\bm k_t}/\absol{\bm P_t}\gtrsim1$, where the HEF and ITMD cross-sections are equal (implying negligible leading-twist saturation corrections), but both different from the CGC one. The figure shows the maximal size of the genuine higher-twist effects, which broaden the dijet angular distribution ($\absol{\bm k_t}/\absol{\bm P_t} \sim \pi - \phi$) and become more visible with heavy nuclear target (large $\hat{c}$).

%%%%%%%%%%%%%%%%%%%%%%%%%%%%%%%%%%%%%%%%%%%%%%%%%%%%%%%%%%%%%%%%%%%%
\section{Summary}{\label{sec:Summary}}
%%%%%%%%%%%%%%%%%%%%%%%%%%%%%%%%%%%%%%%%%%%%%%%%%%%%%%%%%%%%%%%%%%%%

We have compared quantitatively in detail the result of the ITMD formula to that of the CGC formula for forward $q \bar q$ dijet production in $p+p$ and $p+A$ collisions.
We assumed that the typical transverse momentum of a hard jet $\absol{\bm P_t}$ is much bigger than the saturation scale of the target $Q_s$, but considered arbitrary values
of $\absol{\bm k_t}$, the transverse momentum imbalance of the quark-antiquark pair.

First, Sec.\,\ref{sec:Frameworks} has recaptured the differences and similarities between the two frameworks in describing the forward dijet production cross-section.
The ITMD formula \eqref{eq:itmd-qqbar} contains three kinds of leading-twist small-$x$ gluon TMDs, but two of them, $\mathcal{F}_{gg}$ and $\mathcal{F}_{\rm adj}$, are relevant in the large-$N_c$ approximation. At small $\absol{\bm k_t}$ (the TMD regime), the differences between those distributions, see Fig.\,\ref{fig:gtmds}, is the result of an all-order resummation of saturation corrections in $Q_s/\absol{\bm k_t}$, while the hard factors incorporate an all-order resummation of kinematical twists in $\absol{\bm k_t}/\absol{\bm P_t}$, resulting in a proper matching to BFKL at large $\absol{\bm k_t}$ (the HEF or $k_t$-factorization regime). The CGC formula \eqref{eq:cgc-qqbar} involves 2-, 3- and 4-point correlators of Wilson lines; it contains the full ITMD formula and on top resums the genuine higher-twist contributions in $Q_s/\absol{\bm P_t}$.
We should keep in mind that there are these three distinct features embraced as saturation effects in the CGC framework.

Using the Gaussian truncation, however, one can make the two formulae look rather similar: both involve convolutions
of the $q\bar q$ dipole amplitude in momentum space (Eq.\,\eqref{eq:FTofSqq}) with itself and with hard parts.
In the ITMD case \eqref{eq:itmd-large-Nc}, those convolutions are simply the gluon TMDs \eqref{eq:gluontmds-large-Nc}. In the
CGC case they are more involved \eqref{eq:cgc-large-Nc} as they include the genuine higher-twist contributions. Those come from
multi-body correlators (e.g. \eqref{eq:four-point}), but our approximations have allowed us to write them in terms of the function $F$.
The genuine high-twists are suppressed in high-$\absol{\bm p_t}$ dijet production, i.e., $\absol{\bm P_t} \gg Q_s$, in which case the ITMD formula represents
a good approximation to the CGC framework.

In Sec.\,\ref{sec:Results}, we have demonstrated the quantitative difference between the two formulas for forward quark
dijet production by evaluating the azimuthal dijet correlation in $p+p$ and $p+A$ collisions at collider energies.
We have confirmed that the ITMD formula, which interpolates between the TMD and HEF formula, gives the same prediction as
the CGC for the dijets with $\absol{\bm p_t} \sim 40\GeV$ at the LHC energy, where the higher-twist genuine corrections are suppressed.
As $\absol{\bm p_t}$ is decreased, some difference is seen and amount to around $5\textrm{--}15 \%$ for $\absol{\bm p_t} \sim 10\GeV$ at moderate $\phi$ away from
the back-to-back limit. We can regard that amount as the highest estimation of the genuine twist effect for the $q\bar{q}$ dijet correlation,
as well as for the other dijet channels for which those estimations would be more involved.

The nuclear modification factor $R_{pA}$ in $p+A$ collisions shows a dip structure around the back-to-back region of $\phi$ in both the frameworks,
resulting from leading-twist saturation effects in nuclear versus proton targets, and reflecting the intrinsic ${\bm k_t}$ of the gluons, which is of the order of $Q_s$.
For $\absol{\bm p_t} \lesssim 10\GeV$ at moderate $\phi$ away from the back-to-back limit at the LHC, the ITMD gives a suppression while the CGC formula yields an enhancement.
We attribute this difference to a nuclear enhancement of the genuine-twist contributions, i.e., the higher-body multiple scattering effects included in the CGC formula.
The effects are more substantial at lower $\absol{\bm p_t}$ and with the denser nuclear target than higher $\absol{\bm p_t}$ with the dilute one.

When the ITMD formula is used to evaluate the forward dijet production cross-section at moderate $\absol{\bm p_t}$ for the study of gluon saturation, one should be aware of the fact that this framework lacks those genuine twist effects. We note that the studies which are restricted to the TMD regime near $\phi=\pi$, e.g. to isolate the contribution of polarized gluons (relevant when massive quarks are considered \cite{Akcakaya:2012si,Marquet:2017xwy}, for dijets in deep inelastic scattering \cite{Metz:2011wb,Dominguez:2011br,Dumitru:2015gaa,Boer:2016fqd,Dumitru:2018kuw} or for three-particle production \cite{Benic:2017znu,Altinoluk:2018byz,Altinoluk:2020qet}) or to implement a Sudakov resummation \cite{Stasto:2018rci,vanHameren:2019ysa}, are rather safe provided the $\absol{\bm p_t}$'s are not too low.

It would also be interesting to examine whether those effects are experimentally measurable, provided that the gluon TMDs could be determined with good accuracy in other processes. For this purpose, we need to take account of the effects of jet fragmentation and also other effects in jet identification algorithm and efficiency cuts, and so forth. We leave those as future work. 

\acknowledgments

The authors are grateful to Renaud Boussarie, Tolga Altinoluk for useful discussions.
This work was supported
by SAKURA joint research program between France and Japan.
K.W. is supported by Jefferson Science Associates, LLC under U.S. DOE Contract No. DE-AC05-06OR23177. 
K.W. was also supported by U.S. DOE Grant No. DE-FG02-97ER41028 
and the National Science Foundation of China (NSFC) under Grant No. 11575070 when this work was initiated.
HF's work was partly supported by Grant-in-Aid of MEXT 16K05343.
The work of CM is supported by the Agence Nationale de la Recherche under the project ANR-16-CE31-0019-02. 
This work has received funding from the European Union's Horizon 2020 research and innovation programme under grant agreement No. 82409.

\bibliographystyle{JHEP}
\bibliography{BibTexData}

\end{document}